\documentclass[10pt,a4paper]{article}

\usepackage{jheppub}

\usepackage{color}
\usepackage{colordvi}
\usepackage{changes}

\usepackage{epsfig,epsf}
\usepackage{amsmath}
\usepackage{amsthm}
\usepackage{amsfonts}
\usepackage{amssymb}
\usepackage{dsfont}
\usepackage{epstopdf}
\usepackage{multirow}
\usepackage{braket}
\usepackage{tabularx}

\usepackage{tikz}
\usetikzlibrary{arrows.meta,positioning,calc,patterns}

\usepackage{rotating}

\usepackage{marvosym}

\usepackage{todonotes}

\usepackage{subcaption}
\usepackage{array}   
\newcolumntype{C}{>{$}c<{$}}

\usepackage{slashed}

\usepackage[active]{srcltx}

\newcommand{\p}{\partial}
\newcommand{\s}[1]{\slashed{#1}}

\newcommand{\w}[1]{\widetilde{#1}}
\newcommand{\f}[1]{\mathcal{#1}}

\newcommand{\bb}[1]{\mathbb{#1}}
\newcommand{\ms}{\overline{\text{MS}}}

\newcommand \widebar [1] {\overline{#1}}

\def\II{\hbox{{1}\kern-.25em\hbox{l}}}

\title{
The Fate of Ultra-Collinear Modes in On-Shell Massive Sudakov Form Factors}

\author[a,b]{Marvin Schnubel,}
\author[c,d]{Jakob Schoenleber,}
\author[b]{Robert Szafron}
\affiliation[a]{Nikhef, Theory Group, Science Park 105, 1098 XG Amsterdam, The Netherlands}
\affiliation[b]{
   High Energy Theory Group, Physics Department, Brookhaven National Laboratory, Upton, NY 11973, USA}
\affiliation[c]{Physik Department T31, James-Franck-Straße 1, Technische Universität München, D-85748 Garching, Germany}
\affiliation[d]{
   RIKEN BNL Research Center, Brookhaven National Laboratory, Upton, NY 11973, USA}

\emailAdd{m.schnubel@nikhef.nl}
\emailAdd{jakob.schoenleber@tum.de}
\emailAdd{rszafron@bnl.gov}

\abstract{
Individual multi-loop diagrams for the massive Sudakov form factor contain an infinite tower of ultra-collinear momentum regions. We show that, for the on-shell form factor in QCD, these contributions cancel to all orders as a consequence of gauge invariance, so the leading-power SCET$_{\rm II}$ factorization formula is unchanged.
Using the $\eta$ rapidity regulator, we compute the soft function and the massive jet function of the quark and gluon Sudakov form factors through two loops and resum logarithms at NNLL accuracy, including hierarchies of fermion masses.
We also show that with a gauge-boson mass regulator, the infinite tower of modes is truncated and ultra-collinear and ultra-soft modes become manifest and factorize explicitly, providing a direct EFT derivation of the regulated infrared dependence.
}

\begin{document}
\begin{flushright}
{\small
Nikhef 2026-003 \\
TUM-HEP-1598/26
}
\end{flushright}
\maketitle

\section{Introduction}

Method-of-regions studies of the on-shell Sudakov form factor with massive external fermions reveal momentum regions beyond the standard hard, collinear, anti-collinear, and soft scalings. For the two-loop ladder integral, an additional ultra-(anti-)collinear region was identified in Refs.~\cite{Smirnov:1998vk,Smirnov:1999bza}, characterized by a virtuality $m^6/Q^4 \ll m^2$. Moreover, higher-loop graphs generate an infinite cascade of increasingly soft ultra-collinear regions~\cite{Ma:2023hrt,Jaskiewicz:2024xkd,Ma:2025emu,Ma:2026pjx}. The presence of these additional modes potentially invalidates standard factorization; however, as shown by explicit computations in QED, these extra regions cancel in the two-loop QED form factor~\cite{Becher:2007cu,terHoeve:2023ehm}.

Our primary aim is to demonstrate the cancellation of ultra-collinear modes in QCD by virtue of gauge invariance, to all orders.. 
This can be achieved by considering a sequence of EFTs, starting with SCET$_{\rm II}$, followed by an infinite sequence of (equivalent) boosted heavy quark effective theories (bHQETs)~\cite{Fleming:2007qr, Fleming:2007xt}.  This is not intended as a proof of the full factorization theorem\footnote{Concerns have recently been raised about the validity of factorization in the off-shell Sudakov case \cite{Belitsky:2025bez}.}, but to show that the ultra-collinear cascade identified so far does not survive in the on-shell amplitude. Importantly, our argument does not rely on a complete classification of all possible regions, but instead shows that any would-be ultra-collinear contribution is redundant in the presence of gauge invariance and eikonal factorization.

The QCD Sudakov form factor is known from direct calculation at two loops \cite{Bernreuther:2004ih,Gluza:2009yy}. Its phenomenological relevance has been pointed out for hadron and lepton colliders, for example in \cite{deFlorian:2018wcj,Ciafaloni:1999ub,Blumlein:2020jrf,Denner:2000jv}. Partial results for the form factor at three loops in full QCD have been obtained in \cite{Henn:2016kjz,Henn:2016tyf,Ablinger:2017hst,Lee:2018nxa,Lee:2018rgs,Ablinger:2018yae,Blumlein:2019oas,Blumlein:2018tmz,Fael:2022miw,Fael:2022rgm,Fael:2023zqr,Blumlein:2023uuq,Blumlein:2024tjw}, but no complete analytic result exists yet.\footnote{In $\mathcal{N}=4$ super Yang-Mills theories, the Sudakov form factor is already known up to four loops \cite{Lee:2021lkc,Guan:2023gsz,Gehrmann:2011xn}}

The second aim of this work is to provide results and EFT foundations that are relevant to the so-called ``massification'' procedure (also known as IR matching) \cite{Penin:2005eh,Mitov:2006xs,Becher:2007cu,Engel:2018fsb,Wang:2023qbf}. It is the process of reconstructing the massive result from the massless one through the identification
\begin{align}
\frac{1}{\epsilon} \longleftrightarrow \ln m^2 + \text{const,}
\end{align}
with the constant depending on the scheme. 
This formal identification, already observed in \cite{Marciano:1974tv}, can be made rigorous by deriving a factorization theorem: in the limit $m^2\ll Q^2$, the full amplitude splits into a massless hard part and a universal massification ``$Z$-factor''. The latter can be identified as a combination of soft and jet functions. Schematically, one may write
\begin{align}
\underbrace{ 1 + \frac{\alpha}{\pi} \, \ln \frac{m^2}{Q^2} }_{\text{massive amplitude }} = \underbrace{  \left \{ 1 + \frac{\alpha}{\pi} \left ( \frac{1}{\epsilon} + \ln \frac{\mu^2}{Q^2} \right ) \right \} }_{\text{massless amplitude}} \underbrace{  \left \{ 1 - \frac{\alpha}{\pi} \left ( \frac{1}{\epsilon} + \ln \frac{\mu^2}{m^2} \right ) \right \} }_{\text{massification ``$Z$-factor''}} + \, \f O(\alpha^2).
\end{align}
This factorization into single-scale components has multiple advantages. Since the massification factors are universal and process-independent, the massive amplitude can be inferred from the corresponding massless amplitudes, where the calculations are often simpler. Furthermore, large logarithmic corrections in the individual components can be resummed using standard methods, ultimately enabling the resummation of $\ln (m^2/Q^2)$
in the full amplitude.

In section \ref{sec: region cascade general discussion}, we discuss the details of the EFT treatment of the extra modes and their cancellation in the on-shell amplitude. 
In section~\ref{sec:rapreg}, we compute the soft function $S$ and the combined jet function $Z$, which describe the mass dependence of the Sudakov form factor. These functions were calculated in \cite{Becher:2007cu} with a different regulator for rapidity divergences, which are generic to SCET$_{\rm II}$. We use the rapidity regulator defined in \cite{Chiu:2012ir} and provide results for the (anti-)collinear and soft functions to two-loop order. This choice of rapidity regulator enables us to cleanly resum all logarithms $\ln\frac{m^2}{Q^2}$ up to next-to-next-to-leading logarithmic accuracy (NNLL) using both the ordinary renormalization group equations (RGEs) and the rapidity renormalization group equations (RRGEs) \cite{Chiu:2012ir}. 
The single-flavor results coincide with \cite{Hoang:2015vua}. Here we additionally treat an arbitrary number of quark flavors with a mass hierarchy, enabling the resummation of logarithms of quark-mass ratios.

In section~\ref{sec: gluon mass reg}, we apply an alternative IR regularization approach, where IR divergences are regulated by a nonzero gauge-boson mass $m_g$. Strictly speaking, this applies only to QED. We compute all relevant component functions at one loop and show that factors of $\ln m_g$ factorize into ultra-collinear and ultra-soft functions. In section~\ref{sec: abelian exp}, we explicitly show that the well-known exponentiation of IR divergences in QED follows directly.

In section~\ref{sec: gluon FF}, we generalize the results for the quark Sudakov form factor $\bra{q'} \bar \psi \gamma^{\mu} \psi \ket {q}$ to the scalar gluon form factor $\bra{g'} F_{\mu \nu}^A F^{\mu \nu A} \ket g$. Although the gluon is massless, massive fermions do appear in internal fermion loops, giving the gluon $Z$-factor a non-trivial structure. The corresponding soft function and $Z$-factor have been computed in \cite{Wang:2023qbf} with a different rapidity regularization scheme. We argue that the rapidity regulator $\eta$ introduced in \cite{Chiu:2012ir} is more advantageous, as it enables the systematic resummation of rapidity logarithms through rapidity RGEs. 
We extract the gluon soft function and $Z$-factor at NNLO from the results of \cite{Wang:2023qbf}, now employing the $\eta$ rapidity regulator.
We conclude in section~\ref{sec:conclusion}.

\section{Ultra-collinear modes} \label{sec: region cascade general discussion}

In this section, we clarify the status of the ``ultra-collinear'' momentum regions that appear in a method-of-regions analysis of the on-shell Sudakov form factor at higher loops \cite{Smirnov:1999bza}. Our key result is that these regions do not modify the leading-power factorization of the on-shell Dirac form factor. In pure dimensional regularization, the EFT interpretation is a cascade of matching steps between theories with dynamical fluctuations at progressively lower virtualities (bHQET and its generalizations), whose would-be matching functions are scaleless on shell and hence equal to unity.  As a result, the standard SCET$_{\rm II}$ factorization formula remains unchanged at leading power. This cancellation should not be confused with the absence of long-distance sensitivity, since the on-shell form factor is infrared divergent and its factorized components carry nontrivial IR singularities; dimensional regularization simply hides scaleless regions. When an explicit IR regulator is introduced (e.g., a small gluon mass), the same ultra-collinear (and accompanying ultra-soft) modes appear with a physical scale and reproduce the regulated infrared dependence. We proceed by first setting up the SCET$_{\rm II}$ factorization for the on-shell form factor, then exhibiting the ultra-collinear messenger couplings and their decoupling, and finally demonstrating the resulting integrand-level cancellations in explicit multi-loop examples.

\subsection{EFT setup} 
\label{sec:EFT}
We define the Dirac and Pauli form factors of the quark vector current
\begin{align}
  J^\mu(x)=\bar\psi(x)\gamma^\mu\psi(x)
\end{align}
through the on-shell decomposition
\begin{align}\label{eq:SFFdef}
\bra{p_{\bar c}} J^\mu(0) \ket{p_c}
&= \bar u(p_{\bar c})
\left[
\gamma^{\mu} F_1(Q^2,m^2)
+ \frac{i\sigma^{\mu\nu} q_{\nu}}{2m}\, F_2(Q^2,m^2)
\right]
u(p_c),
\end{align}
with \(q\equiv p_{\bar c}-p_c\), \(Q^2\equiv -q^2\), and on-shell external legs \(p_c^2=p_{\bar c}^2=m^2\). We consider the Sudakov regime $Q^2\gg m^2$ and expand in the small parameter
\begin{align}
\lambda \equiv \frac{m}{Q}\ll 1.
\end{align}
Since the Pauli term multiplies a helicity-flip structure, \(F_2\) is power suppressed at large \(Q\). We restrict the discussion to leading power effects in this work and therefore focus on \(F_1\).

In QED,~\eqref{eq:SFFdef} defines a gauge-invariant on-shell form factor of the renormalized vertex function. Current conservation implies that the vector current does not require independent UV renormalization.\footnote{A careful discussion of the renormalization of the electromagnetic current (including operator mixing with EOM terms in \(\overline{\rm MS}\)) can be found in \cite{Collins:2005nj}.}
After renormalizing the fields, coupling, and mass, any remaining poles in the on-shell form factors are infrared in origin.
With a massless photon, the form factor is IR divergent and must be defined with an IR regulator; the associated singularities cancel in inclusive observables (or can be absorbed into dressed asymptotic states).

In QCD, quarks are not asymptotic states, but~\eqref{eq:SFFdef} still provides the standard perturbative (partonic) definition obtained by LSZ amputation of on-shell quark external legs, where \(m\) denotes the corresponding pole-mass parameter.

In the first part of the paper, we consider $SU(N_c)$ Yang-Mills theory with a single massive quark of mass $m$,
\begin{align}\label{eq:QCDL}
\mathcal{L}
= \bar \psi\,(i\slashed{\partial}+g\slashed{A}-m)\psi
-\frac14 F_{\mu\nu}^A F^{\mu\nu A},
\end{align}
with gauge-fixing and ghost terms understood. We use Feynman gauge throughout.
For brevity, we refer to this theory as ``QCD'' and to $\psi$ and $A_\mu^A$ as the quark and gluon.
The QED limit follows by the usual abelianization of the color factors, and the extension to several quark flavors is discussed in section~\ref{sec: more quarks}.

We employ dimensional regularization, $d=4-2\epsilon$ for UV and IR divergences unless specified otherwise. For subsequent comparison with the physical QED limit, we use the pole mass scheme. At one loop, the bare mass $m_0$ and pole mass $m$ are related by
\begin{align} \label{eq: m renorm}
m_0^2 = m^2 \Bigg \{ 1 + \frac{\alpha_0 C_F}{4\pi} \left ( \frac{\mu^2}{m^2} \right )^{\epsilon} (4\pi)^{\epsilon} \Gamma(1+\epsilon) \frac{-6+4\epsilon}{\epsilon(1-2\epsilon)} + \f O(\alpha_0^2) \Bigg \},
\end{align}
where the bare coupling constant is $\alpha_0=g_0^2/(4\pi)$. $C_F$ is the quadratic Casimir of the $SU(N_{c})$ fundamental representation.

In the center-of-mass frame, we decompose momenta into light-cone components as $p^\mu = (n_+ p)\, n_-^\mu/2 + (n_- p)\, n_+^\mu/2 + p_\perp^\mu$ with $n_\pm^2=0$ and $n_+\!\cdot n_-=2$.
The external momenta scale as\footnote{ When denoting scaling, we order the comments as $(n_+ p,p_\perp,n_-p)$.}
\begin{align}
p_c^\mu \sim Q(1,\lambda,\lambda^2), \qquad p_{\bar c}^\mu \sim Q(\lambda^2,\lambda,1),
\end{align}
and soft momenta scale as
\begin{align} \label{eq: soft def}
p_s^\mu \sim Q(\lambda,\lambda,\lambda).
\end{align}

The hierarchy $Q^2 \gg Qm \gg m^2$ admits the standard two-step construction
${\rm QCD}\to{\rm SCET}_{\rm I}\to{\rm SCET}_{\rm II}$ \cite{Beneke:2003pa}.
The intermediate ${\rm SCET}_{\rm I}$~~\cite{Bauer:2000yr,Bauer:2001yt,Bauer:2002nz,Beneke:2002ph,Beneke:2002ni} contains hard-collinear modes, which contribute non-trivial matching factors only in the presence of soft external particles \cite{Hill:2002vw,Bonocore:2015esa,Beneke:2003pa}, commonly beyond leading power \cite{Beneke:2018gvs,Moult:2018jjd,Moult:2019mog,Beneke:2019mua,Beneke:2019oqx,Beneke:2020ibj,Liu:2020ydl,Liu:2021mac,Beneke:2022obx,Liu:2022ajh,vanBeekveld:2023liw}. We tacitly skip the discussion of the intermediate effective theory, as it follows standard procedures.

In what follows, we work directly with the final ${\rm SCET}_{\rm II}$ description, which is valid for virtualities of order $m^2$.
At leading power, the SCET$_{\rm II}$ Lagrangian splits into independent collinear, anti-collinear, and soft sectors,
\begin{align}
\f L_{\rm SCET_{\rm II}} = \f L_c(\xi_c , A_c) + \f L_{\bar c}(\xi_{\bar c}, A_{\bar c}) + \f L_s(q_s, A_s)\, .
\end{align}
where $\f L_s$ is the Lagrangian in eq.~\eqref{eq:QCDL} restricted to soft modes, and the collinear Lagrangian reads \cite{Leibovich:2003jd,Chay:2005ck}
\begin{align} \label{eq: Lc}
\f L_c = \bar \xi_c \left [ in_- D_c  + (i \s D_{c\perp} - m ) \frac{1}{in_+ D_c} (i \s D_{c\perp} + m )  \right ] \frac{\s n_+}{2} \xi_c - \frac{1}{4} F_{c\mu \nu}^a F_c^{\mu \nu a},
\end{align}
subject to the standard definitions \cite{Beneke:2002ni,Beneke:2003pa}. The mass, field, and coupling renormalization factors in SCET are the same as in QCD \cite{Chay:2005ck}, which is consistent with the SCET no-renormalization theorem~\cite{Beneke:2002ph}.

The soft Wilson lines induced in the currents by integrating out the hard-collinear virtuality $mQ$ are defined as
\begin{align} \label{eq: soft wline 1}
S_{\pm}(x) = \mathcal{ P} \exp \left [ ig \int_{-\infty}^0 ds \, n_{\pm}\!\cdot A_s(x + s n_{\pm} ) \right ] .
\end{align}
Here $\mathcal{P}$ denotes path ordering, which can be omitted in abelian gauge theories. An appropriate $i\epsilon$ prescription is implied to define the integral, as dictated by the underlying QCD dynamics. Further details can be found in appendix A of~\cite{Beneke:2019slt}.

The result is then the following leading power operator level matching relation (all operators evaluated at position $x=0$)
\begin{align}
\label{eq:Jmatching}
J^\mu(0) =   C(Q^2,\mu) \bar \xi_{\bar c} W_{\bar c} S_-^{\dagger} \gamma_{\perp}^{\mu} S_+ W_c^{\dagger} \xi_c +\mathcal{O}(\lambda). 
\end{align}
The collinear and anti-collinear Wilson lines in the fundamental representation are
\begin{align} \label{eq: coll wline 1}
W_c(x) &= \mathcal{ P} \exp \left ( i g \int_{-\infty}^0 ds \, n_+ \cdot A_c(x+sn_+) \right ),
\\ \label{eq: acoll wline 1}
W_{\bar c}(x) &= \mathcal{ P} \exp \left ( ig \int_{-\infty}^0 ds \, n_- \cdot A_{\bar c}(x + sn_-) \right ).
\end{align}
The matching equation~\eqref{eq:Jmatching} yields the familiar factorization formula
\begin{align}\label{eq:fact1}
F_1(Q^2,m^2) = C(Q^2,\mu)\, Z^{1/2}_c(m^2,\mu,\nu)\, Z^{1/2}_{\bar c}(m^2,\mu,\nu)\, S(m^2,\mu,\nu)
+ \mathcal O(\lambda^2)\,,
\end{align}
where $C$ is the hard matching coefficient (equal to the massless quark on-shell form factor after IR subtraction, known to four loops \cite{Lee:2022nhh}) and the dependence on the renormalization scale $\mu$ and rapidity scale $\nu$ is shown explicitly. The collinear functions are defined by single-particle matrix elements\footnote{(Anti-)collinear spinors are $u_c=\frac{\slashed n_- \slashed n_+}{4}u(p_c)$ and $u_{\bar c}=\frac{\slashed n_+ \slashed n_-}{4}u(p_{\bar c})$.}
\begin{align}
\sqrt{Z_c}\,u_c=\langle 0|W_c^\dagger \xi_c|p_c\rangle\,,\qquad
\sqrt{Z_{\bar c}}\,u_{\bar c}=\langle 0|W_{\bar c}^\dagger \xi_{\bar c}|p_{\bar c}\rangle\,,
\end{align}
and the soft function is the vacuum matrix element of soft Wilson lines
\begin{align}\label{eq:Sdef}
S=\frac{1}{N_c}\,\mathrm{tr}\,\langle 0|S_-^\dagger S_+|0\rangle\,.
\end{align}
Because soft and (anti-)collinear modes share the same virtuality $\mathcal{O}(m^2)$ in SCET$_{\rm II}$, the separate factors $Z_{c,\bar c}$ and $S$ are rapidity divergent and require regularization together with the appropriate overlap (zero-bin) subtraction. Only the product in~\eqref{eq:fact1} is rapidity regulator independent. We provide more details in section~\ref{sec:rapreg}.

As shown in \cite{Smirnov:1999bza}, in the two-loop ladder diagram contributing to $F_1$, the sum of the standard hard, (anti-)collinear, and soft regions does not reproduce the correct asymptotic expansion at leading power in $\lambda$. The missing contribution is captured by an additional ultra-collinear region with momentum scaling
\begin{align}\label{eq:c+uc}
p_{uc}^{\mu} \sim Q(\lambda^2, \lambda^3, \lambda^4), \qquad
p_{\widebar{uc}}^{\mu} \sim Q(\lambda^4, \lambda^3, \lambda^2)\,.
\end{align}
In EFT terms, the appearance of \eqref{eq:c+uc} signals sensitivity to a lower off-shellness scale below $m^2$: when an on-shell collinear line emits an ultra-collinear momentum, its virtuality changes only by a parametrically smaller amount; see eq.~\eqref{eq:puc_add} below. To keep each factorized ingredient single-scale, we therefore match SCET$_{\rm II}$ onto a lower-energy theory with dynamical ultra-collinear modes \cite{Pietrulewicz:2014qza,Hoang:2015iva}. Now, the objects $Z_{c,\bar c}$ and $S$ in \eqref{eq:fact1} can be viewed as Wilson coefficients rather than as matrix elements. This viewpoint cleanly separates UV poles associated with matching from infrared poles and aligns with the overlap/soft-subtraction logic used at the amplitude level \cite{Collins:1999dz,Beneke:2019slt}.

For an on-shell massive collinear momentum $p_c^2=m^2$, $p_{uc}$ acts as a residual fluctuation on top of a fixed large momentum component,
\begin{align}\label{eq:puc_add}
(p_c+p_{uc})^2-m^2 = 2\,p_c\!\cdot p_{uc}+\mathcal O(p_{uc}^2)\,,
\end{align}
so that the interaction admits an HQET-type expansion. By contrast, for an anti-collinear momentum,
\begin{align}\label{eq:pubc_add}
(p_{\bar c} + p_{uc})^2 - m^2 \sim n_-p_{\bar c}\, n_+ p_{uc}\,,
\end{align}
the ultra-collinear transfer induces a virtuality of order $m^2$ and thus does not drive the $\bar c$ sector off shell. In this sense, ultra-collinear modes can act as messenger fields between the two collinear sectors \cite{Becher:2003qh}; the description is symmetric under $c\leftrightarrow \bar c$ (with $p_{\widebar{uc}}$ coupling analogously). Such messenger modes are often referred to as soft-collinear \cite{Becher:2003qh,Pietrulewicz:2014qza,Hoang:2015iva} because they mediate interactions between otherwise decoupled sectors, in analogy to ultra-soft modes in SCET$_{\rm I}$.

\begin{figure}
    \centering
\begin{tikzpicture}[x=1.45cm,y=1.45cm,>=Latex,font=\small]


\definecolor{gridCol}{HTML}{DDE7E1}
\definecolor{boxGreen}{HTML}{80bff9}
\definecolor{boxOrange}{HTML}{741b47}

\definecolor{v0}{HTML}{08306B}
\definecolor{v1}{HTML}{0B2C6B}
\definecolor{v2}{HTML}{0B4FA2}
\definecolor{v4}{HTML}{3B87C8}
\definecolor{v6}{HTML}{A8D5EE}

\definecolor{mOrange}{HTML}{741b47}

\tikzset{axis/.style={line width=0.8pt, draw=black}}
\tikzset{grid/.style={gridCol, very thin}}

\tikzset{lblTop/.style={draw=black, rounded corners, fill=white,
                        inner sep=3pt, align=left, text=black}}
\tikzset{lblMassless/.style={draw=black, rounded corners, fill=boxGreen!18,
                             inner sep=3pt, align=left, text=black}}
\tikzset{lblMassive/.style={draw=black, rounded corners, fill=boxOrange!22,
                            inner sep=3pt, align=left, text=black}}

\tikzset{diag1/.style={draw=v1, dashed, line width=1.0pt}}
\tikzset{diag2/.style={draw=v2, line width=1.5pt}}
\tikzset{diag4/.style={draw=v4, line width=1.5pt}}
\tikzset{diag6/.style={draw=v6, line width=1.5pt}}

\tikzset{band0/.style={fill=v0, opacity=0.14}}
\tikzset{band2/.style={fill=v2, opacity=0.14}}
\tikzset{band4/.style={fill=v4, opacity=0.12}}
\tikzset{band6/.style={fill=v6, opacity=0.10}}

\tikzset{L2/.style={draw=mOrange, line width=3.0pt,
                    pattern=north east lines, pattern color=mOrange}}
\tikzset{L4/.style={draw=mOrange, line width=3.0pt,
                    pattern=north east lines, pattern color=mOrange}}

\tikzset{guide/.style={draw=mOrange, dotted, line width=1.5pt}}

\tikzset{mode/.style={circle, draw=black, fill=white, line width=0.8pt,
                      minimum size=18pt, inner sep=0pt, font=\small\bfseries}}
\tikzset{modeLabel/.style={font=\footnotesize, text=black}}

\path[use as bounding box] (-0.55,-0.55) rectangle (7.20,7.05);

\draw[axis,->] (-0.1,-0.1) -- (6.6,-0.1) node[below right]
  {$n_{+}k/Q \sim \lambda^{a}$};
\draw[axis,->] (-0.1,-0.1) -- (-0.1,6.2) node[above ]
  {$n_{-}k/Q \sim \lambda^{b}$};

\draw[grid, step=1] (0,0) grid (6.2,6.0);

\foreach \t in {0,1,2,3,4,5,6} {
  \node[text=black] at (\t,-0.35) {$\lambda^\t$};
  \node[text=black] at (-0.35,\t) {$\lambda^\t$};
}


\fill[band0] (0,0) -- (0,1) -- (1,0) -- cycle;

\fill[band2] (0,1) -- (0,2) -- (2,0) -- (1,0) -- cycle;

\fill[band4] (0,2) -- (0,4) -- (4,0) -- (2,0) -- cycle;

\fill[band6] (0,4) -- (0,6) -- (6,0) -- (4,0) -- cycle;

\draw[diag1] (0,1) -- (1,0);
\draw[diag2] (0,2) -- (2,0);
\draw[diag4] (0,4) -- (4,0);

\draw[diag6] (0,6) -- (6,0);

\draw[guide] (0,2) -- (4,6);
\draw[guide] (0,0) -- (0,2);

\draw[L2] (0,2) -- (6.2,2);
\draw[L2] (0,2) -- (0,6.0);

\draw[L4] (2,4) -- (6.2,4);
\draw[L4] (2,4) -- (2,6.0);

\node[lblMassless, anchor=south west] at (0.62,0.12)
  {{$k^2\sim Q^2\lambda^{1}$}};
\node[lblMassless, anchor=south west] at (1.25,0.62)
  {{$k^2\sim Q^2\lambda^{2}$}};
\node[lblMassless, anchor=south west] at (2.55,1.2)
  {{$k^2\sim Q^2\lambda^{4}$}};
\node[lblMassless, anchor=south west] at (4.55,1.35)
  {{$k^2\sim Q^2\lambda^{6}$}};

\node[lblMassive, anchor=west] at (6.25,2.05)
  {$\Delta_c \sim Q^2\lambda^{2}$};
\node[lblMassive, anchor=west] at (6.25,4.05)
  {$\Delta_c \sim Q^2\lambda^{4}$};

\node[lblTop, anchor=north west] (H) at (1.5,6.25) {%
\begin{minipage}{6.95cm}
\centering
$(n_+k,\,k_\perp,\,n_-k)\sim Q(\lambda^{a},\,\lambda^{(a+b)/2},\,\lambda^{b})$

\vspace{2mm}

\hspace*{\fill}%
\begin{minipage}{0.48\linewidth}
\centering
\fcolorbox{black}{boxGreen!18}{%
\begin{minipage}{0.95\linewidth}
\centering
\textbf{Massless Virtuality}\\
$k^2\sim Q^2\lambda^{a+b}$
\end{minipage}}
\end{minipage}
\hspace*{\fill}
\begin{minipage}{0.48\linewidth}
\centering
\fcolorbox{black}{boxOrange!22}{
\begin{minipage}{0.95\linewidth}
\centering
\textbf{Massive Virtuality}\\
$\Delta_c \sim Q^2\lambda^{\min(b,\,a+2)}$
\end{minipage}}
\end{minipage}
\hspace*{\fill}
\end{minipage}
};

\node[mode] at (0,0) {$h$};

\node[mode] at (0,1) {$hc$};

\node[mode] at (0,2) {$c$};
\node[mode] at (2,0) {$\overline{c}$};

\node[mode] at (1,1) {$s$};
\node[mode] at (2,2) {$us$};
\node[mode] at (3,3) {$us_1$};

\node[mode] at (2,4) {$uc$};
\node[mode] at (4,2) {$\overline{uc}$};

\end{tikzpicture}
\caption{   \label{fig:powercount}
    Power-counting diagram for a momentum $k$ in light-cone coordinates, with
$n_+\!\cdot k/Q\sim\lambda^{a}$ and $n_-\!\cdot k/Q\sim\lambda^{b}$ (hence
$k_\perp/Q\sim\lambda^{(a+b)/2}$), where $\lambda\equiv m/Q$.
Diagonal lines $a+b=\text{const}$ (and the shaded bands between them) correspond to the
massless virtuality $k^2\sim Q^2\lambda^{a+b}$.
For an on-shell massive collinear line, the induced off-shellness
$\Delta_c\equiv (p_c+k)^2-m^2$ scales as $\Delta_c\sim Q^2\lambda^{\min(b,\,a+2)}$; the contours highlight $\Delta_c\sim m^2$ and $\Delta_c\sim m^4/Q^2$.
Marked points indicate the hard, hard-collinear, (anti-)collinear, soft, ultra-soft, and
ultra-collinear modes used in the EFT description. }
\end{figure}
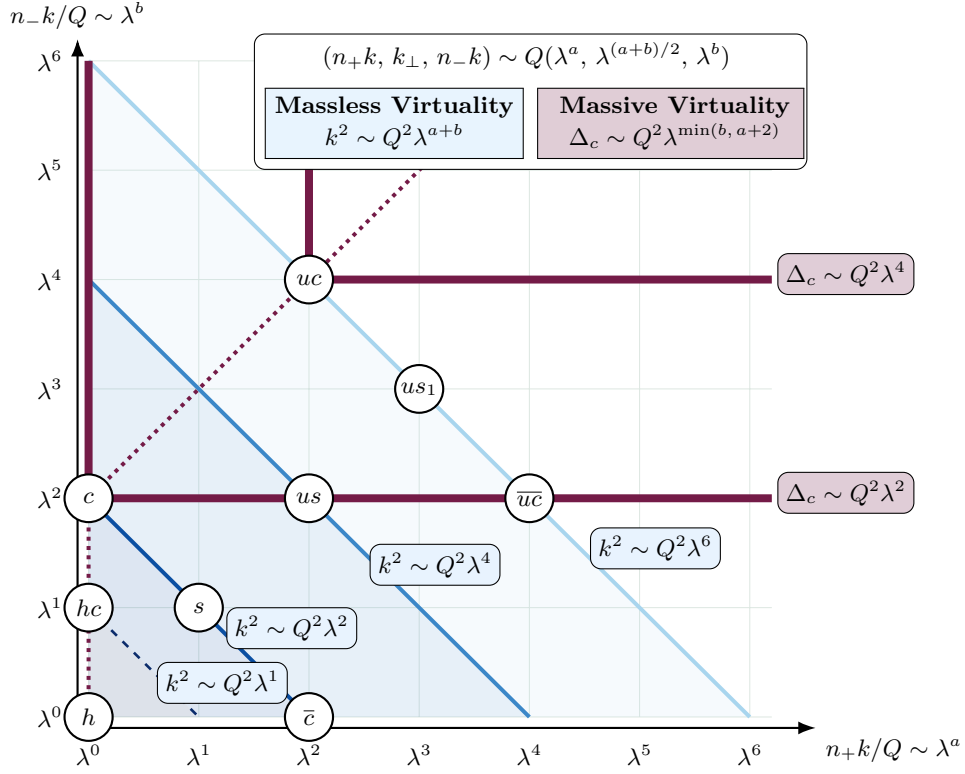

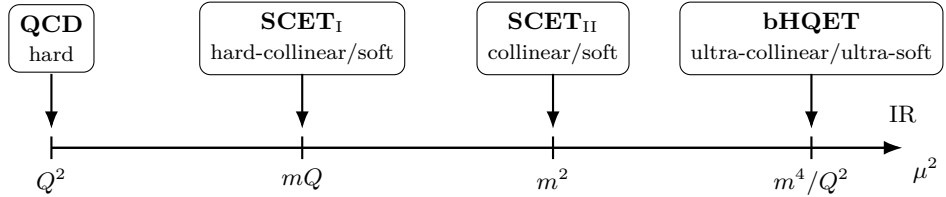
\begin{figure}[t]
\centering
\begin{tikzpicture}[x=1cm,y=1cm,>=Latex,font=\small]

\tikzset{
  scaleaxis/.style={line width=0.9pt, draw=black},
  tick/.style={line width=0.8pt, draw=black},
  eftbox/.style={draw=black, rounded corners, fill=white, inner sep=4pt, align=center},
  lbl/.style={text=black}
}

\def\xQ{0.0}
\def\xmQ{3.3}
\def\xmTwo{6.6}
\def\xIR{10.0}

\draw[scaleaxis,->] (\xQ,-0.2) -- (\xIR+1.2,-0.2) node[below right] {$\mu^2$};

\draw[tick] (\xQ,-0.35) -- (\xQ,-0.05);
\node[lbl, below] at (\xQ,-0.35) {$Q^2$};

\draw[tick] (\xmQ,-0.35) -- (\xmQ,-0.05);
\node[lbl, below] at (\xmQ,-0.35) {$mQ$};

\draw[tick] (\xmTwo,-0.35) -- (\xmTwo,-0.05);
\node[lbl, below] at (\xmTwo,-0.35) {$m^2$};

\draw[tick] (\xIR,-0.35) -- (\xIR,-0.05);
\node[lbl, below] at (\xIR,-0.35) {$m^4/Q^2$};

\node[lbl] at (\xIR+1.2,0.25) {IR};

\node[eftbox] at (\xQ,1.25) {\textbf{QCD}\\[-1pt]\footnotesize hard};
\node[eftbox] at (\xmQ,1.25) {\textbf{SCET$_{\mathrm{I}}$}\\[-1pt]\footnotesize hard-collinear/soft};
\node[eftbox] at (\xmTwo,1.25) {\textbf{SCET$_{\mathrm{II}}$}\\[-1pt]\footnotesize collinear/soft};
\node[eftbox] at (\xIR,1.25) {\textbf{bHQET}\\[-1pt]\footnotesize ultra-collinear/ultra-soft};

\foreach \x in {\xQ,\xmQ,\xmTwo,\xIR} {
  \draw[->, line width=0.8pt] (\x,0.78) -- (\x,0.05);
}

\end{tikzpicture}
\caption{
Chain of EFTs for \emph{massive} energetic particles/jets, organized by the relevant virtuality scale: QCD (hard) at $\mu^2\!\sim\!Q^2$, matched onto SCET$_{\mathrm{I}}$ (hard-collinear) at $\mu^2\!\sim\!mQ$, then SCET$_{\mathrm{II}}$ (collinear/soft) at $\mu^2\!\sim\!m^2$, and finally onto bHQET (ultra-collinear) for scales below $m^2$.
}
\label{fig:eft_chain}
\end{figure}

The partition of the phase space, valid up to two loops, for a generic momentum into modes and their power counting in light-cone coordinates is summarized in
figure~\ref{fig:powercount}. The resulting EFT cascade implied by the hierarchy
$Q^2 \gg Qm \gg m^2$ is shown in figure~\ref{fig:eft_chain}. 

The existence of messenger (ultra-collinear) modes implies that the SCET$_{\rm II}$ description based only on soft and (anti-)collinear fields is not closed: an ultra-collinear gluon can couple to the opposite collinear sector without driving it off shell. Concretely, writing the anti-collinear covariant derivative as
\begin{align}
iD_{\bar c}^\mu(x) &\equiv i\partial^\mu + g\,A_{\bar c}^\mu(x)\,,
\end{align}
the leading-power $\bar c$ Lagrangian requires the replacement
\begin{align}\label{eq:Dbarc_uc}
iD_{\bar c}^\mu(x) \;\longrightarrow\; iD_{\bar c+uc}^\mu(x) &\equiv i\partial^\mu + g\,A_{\bar c}^\mu(x)+  g\,n_+ \cdot A_{uc}(x_+)\frac{n_-^\mu}{2}\,,
\qquad
x_+^\mu \equiv \frac{n_-\!\cdot x}{2}\,n_+^\mu \,.
\end{align}
Here, the ultra-collinear field is multipole expanded in the $\bar c$ sector,
$A_{uc}^\mu(x)=A_{uc}^\mu(x_+)+\mathcal O(\lambda^2)$,
since its $k_\perp$ and $n_-k$ components are parametrically too small to resolve $x_\perp$ and $n_+x$ variations. Multipole expansion is implemented following the method of \cite{Beneke:2002ni}.
By $c\leftrightarrow \bar c$ exchange one obtains the analogous coupling of $\widebar{uc}$ modes to the collinear sector,
\begin{align}\label{eq:Dc_uc}
iD_{c}^\mu(x) \;\longrightarrow\; iD_{c+\overline{uc}}^\mu(x) &\equiv i\partial^\mu + g\,A_{ c}^\mu(x)+  g\,n_- \cdot A_{\overline{uc}}(x_-)\frac{n_+^\mu}{2}\,,
\qquad
x_-^\mu \equiv \frac{n_+\!\cdot x}{2}\,n_-^\mu \,.
\end{align}

Before integrating out modes with virtuality $\mathcal{O}(Q^2 \lambda^2)$, we must decouple the ultra-collinear gluons from the anti-collinear modes and the ultra-anti-collinear gluons from the collinear modes in the SCET$_{\rm II}$ Lagrangian. This can be done by performing a decoupling transformation 
\begin{align}\label{eq:uc_decouple_fields}
\xi_c(x) &= W_{\widebar{uc}}(x_-)\,\xi_c^{(0)}(x),
&\quad
iD_c^\mu(x) &= W_{\overline{uc}}(x_-)\, iD_c^{(0)\mu}(x)\, W_{\overline{uc}}^\dagger(x_-),
\\
\xi_{\bar c}(x) &= W_{uc}(x_+)\,\xi_{\bar c}^{(0)}(x),
&\quad
iD_{\bar c}^\mu(x) &= W_{uc}(x_+)\, iD_{\bar c}^{(0)\mu}(x)\, W_{uc}^\dagger(x_+),
\end{align}
where 
\begin{align}
W_{\widebar{uc}}(x) &= \mathcal{ P} \exp \left [ ig \int_{-\infty}^0 ds\, n_- \cdot A_{\widebar{uc}}(x+sn_-)  \right ],
\\
W_{uc}(x) &= \mathcal{ P} \exp \left [ ig \int_{-\infty}^0 ds\, n_+ \cdot A_{uc}(x+sn_+)  \right ].
\end{align}
The ultra-collinear Wilson lines appear in the currents, in analogy to soft Wilson lines after integrating out the hard-collinear virtuality. Subsequently, we find the current
\begin{align}
J^\mu(0) = C(Q^2) \bar \xi^{(0)}_{\bar c} W_{\bar c} W_{uc}^{\dagger} S_-^{\dagger} \gamma_{\perp}^{\mu} S_+  W_{\widebar{uc}} W_c^{\dagger} \xi^{(0)}_c + \mathcal{O}(\lambda). 
\end{align}

We can now proceed to expand the Lagrangian. We define the velocity four-vectors as
\begin{align}
v_{\pm}^{\mu} = \frac{Q}{m}\frac{ n_{\pm}^{\mu}}{2} + \frac{m}{Q} \frac{n_{\mp}^{\mu}}{2}.
\end{align}
The appropriate low-energy EFT is boosted heavy quark effective theory (bHQET)
\cite{Fleming:2007xt,Fleming:2007qr,Dai:2021mxb}. The boosted heavy-quark fields are defined as
\begin{align}\label{eq: huc def}
h_{uc}(x) = \sqrt{\frac{2}{n_+ v_-}}\,e^{im v_-\!\cdot x}\,\xi^{(0)}_c(x), \qquad
h_{\widebar{uc}}(x) = \sqrt{\frac{2}{n_-v_+}}\,e^{im v_+\!\cdot x}\,\xi^{(0)}_{\bar c}(x).
\end{align}
The parameter $m$ specifies the large kinematic component in the momentum split $p^\mu = m v_{\pm}^\mu + k^\mu$.
In perturbation theory, one may identify $m$ with the on-shell (pole) mass, but in QCD, the pole mass is intrinsically ambiguous at
$\mathcal{O}(\Lambda_{\rm QCD})$; this ambiguity is compensated by an equal and opposite scheme dependence in bHQET matrix elements
(or, equivalently, by a residual mass term in the EFT) \cite{Beneke:1994sw,Bigi:1994em}.
Accordingly, one is free to use any short-distance mass scheme (e.g.\ $\overline{\rm MS}$), with the corresponding matching carried out
perturbatively; our convention is fixed by the renormalization prescription in \eqref{eq: m renorm}.
For QED, where an on-shell mass is physically meaningful, the on-shell scheme provides the most direct correspondence to the physical case.\footnote{We note that the on-shell scheme contains ``hidden'' large corrections that require resummation in the high-energy region when comparing different schemes, cf.~\cite{Jaskiewicz:2024xkd}.}

The corresponding Lagrangians for the ultra-(anti-)collinear fields read \cite{Fleming:2007qr,Fleming:2007xt,Beneke:2023nmj}: 
\begin{align}
\f L_{uc} = \bar h_{uc} i v_- \cdot D_{uc} \frac{\s n_+}{2} h_{uc} - \frac{1}{4} F_{uc \mu \nu}^A F_{uc}^{\mu \nu A}, \qquad \f L_{\widebar{uc}} = \bar h_{\widebar{uc}} iv_+ \cdot D_{\widebar{uc}} \frac{\s n_- }{2 } h_{\widebar{uc}} - \frac{1}{4} F_{\widebar{uc} \mu \nu}^A F_{\widebar{ uc }}^{\mu \nu A},
\end{align}
where $F_{uc}^{\mu \nu}(F_{\widebar{uc}}^{\mu \nu})$ is the gluon field strength tensor of the ultra-(anti-)collinear fields and $iD_{uc}^{\mu} = i \p^{\mu} + g A_{uc}^{\mu}, \, iD_{\widebar{ uc }}^{\mu} = i \p^{\mu} + gA_{\widebar{ uc }}^{\mu}$. The complete Lagrangian describing modes with virtuality up to $\lambda^4 Q^2$ reads
\begin{align} \label{eq: LbHQET}
\f L_{\rm bHQET} = \f L_{uc} + \f L_{\widebar{uc}} + \f L_{us_1}.
\end{align}
where $\f L_{us_1}$ is the Lagrangian in eq.~\eqref{eq:QCDL} restricted to ultra-soft modes. 

We note that the \emph{standard} ultra-soft modes with homogeneous scaling
\begin{equation}
p_{us} \sim (\lambda^2,\lambda^2,\lambda^2)\,,
\end{equation}
correspond to the usual SCET$_{\mathrm{I}}$ soft scaling. Such modes can be viewed as arising from the interaction of ultra-collinear gluons belonging to different sectors, in the sense that their combined scaling satisfies
\begin{equation}
p_{uc} + p_{\widebar{uc}} \sim p_{us}\,,
\end{equation}
at the level of power counting.
However, these are \emph{not} the ultra-soft modes that become relevant once an explicit infrared regulator, such as a gluon mass, is introduced. In that case, the infrared structure resolves a hierarchy of lower-virtuality modes, and one must consider in particular
\begin{equation}
p_{us_1} \sim (\lambda^3,\lambda^3,\lambda^3)\,,
\end{equation}
which correspond to the first non-trivial level of an infinite tower of ultra-soft modes. 
When interacting with ultra-collinear modes, $p_{us_1}$ induces a parametrically larger off-shellness, in analogy with SCET$_{\mathrm{II}}$, where $p_c + p_s \sim p_{hc}$. In the present case, $p_{uc} + p_{us_1}$ probes a higher virtuality than $p_{uc}$ alone, and consequently, the modes in different sectors are decoupled in bHQET

Unless we introduce a rapidity regulator that induces an ultra-soft scale, the ultra-soft function does not contribute to the factorization formula, as each individual ultra-soft contribution is scaleless. We note that, unlike for the soft scale, where there exists a natural mass scale $m^2$ that indeed contributes due to bubble diagrams with a massive fermion, there is no natural ultra-soft scale in the problem, as we do not have any massive modes below $m$ in perturbation theory.

The operator matching relation from SCET$_{\rm II}$ to bHQET is given by
\begin{align} \label{eq: matching 2}
\bar \xi^{(0)}_{\bar c} W_{\bar c}  W_{uc}^{\dagger} S_-^{\dagger} \gamma_{\perp}^{\mu} S_+  W_{\widebar{uc}} W_c^{\dagger} \xi^{(0)}_c = Z^{1/2}_c(m^2)\, Z^{1/2}_{\bar c}(m^2) S(m^2) \; \bar h_{\widebar{uc}} W_{\widebar{uc}}\gamma_{\perp}^{\mu}  W_{uc}^{\dagger} h_{uc} + \mathcal{O}(\lambda).
\end{align}
After integrating out virtualities of order $m^2$, i.e., soft and collinear modes, the previously identified $Z^{1/2}_{c}$, $Z^{1/2}_{\bar c}$, and $S$ matrix elements should be interpreted as matching coefficients for the matching SCET$_{\rm II}$$\to$ bHQET.

Analogously to the collinear functions, we define the ultra-collinear functions
\begin{align}\label{eq:ZUCdef}
\sqrt{\mathfrak Z_{1c}} \, u_c = \bra 0 W_{uc}^{\dagger} h_{uc} \ket{p_c}, \qquad 
\sqrt{\mathfrak Z_{1\bar{c}}} \, \bar u_{\bar{c}} = \bra {p_{\bar c}} \bar h_{\widebar{uc}} W_{\widebar{uc}} \ket{0}.
\end{align}
We note that it is possible to work with a single function, as both $\mathfrak Z_{1c}$ and $\mathfrak Z_{1\bar{c}}$ are equal for a symmetric rapidity regulator, and we define 
\begin{align}
    \mathfrak Z_{1c} =\mathfrak Z_{1\bar{c}} \equiv \mathfrak Z_{1} . 
\end{align}
One readily finds that in the on-shell limit, the loop corrections to $\mathfrak Z_1$ vanish because they are scaleless, i.e., $\mathfrak Z_1 \equiv 1$ to all orders in perturbation theory. This shows that the ultra-collinear modes cancel and do not enter bare factorization in eq. \eqref{eq:fact1}, confirming the findings of \cite{terHoeve:2023ehm} at leading power in QED and generalizing them to all orders in QCD. This fact has an intuitive explanation: there is no physical scale $\mu^2\sim \lambda^4 Q^2 = m^4/Q^2$ that is relevant for the on-shell form factor. 
Even at the level of an infrared-divergent amplitude, the singular structure is completely captured by soft and collinear modes. Gauge invariance enforces eikonal interactions and their representation in terms of Wilson lines, which eliminates any sensitivity to an independent ultra-collinear scale and renders this region redundant in the factorized description.

Beyond two loops, we need to consider the ultra$^j$-collinear modes with scalings
\begin{align} \label{eq: j modes}
p_{c_j} \sim Q(1, \lambda, \lambda^2) \lambda^{2j}, \qquad p_{\bar c_j} \sim Q(\lambda^2, \lambda, 1) \lambda^{2j}, \qquad {\rm for} \qquad j \in \bb N,
\end{align}
with (ultra-)collinear modes corresponding to $j=0$ ($j=1$).
These modes will generally appear in higher loop diagrams, as we will show in the next section. We then expand the sum of collinear and ultra$^j$-collinear momenta in propagators as
\begin{align}
(p_c + p_{c_j})^2 - m^2 \sim 2 p_c \cdot p_{c_j}, \qquad (p_c + p_{\bar c_j})^2 - m^2\sim n_+ \cdot p_c \, n_- \cdot p_{\bar c_j},
\end{align}
which is equivalent to the expansion when the first ultra-collinear mode is considered \eqref{eq:puc_add}. Hence, the correct EFT to describe this tower of lower virtuality ultra-collinear modes is a tower of stacked but equivalent bHQETs. Denoting the $j$-th ultra-collinear EFT as bHQET$_j$, we can write their Lagrangian as 
\begin{align}
\f L_{{\rm bHQET}_j} = \bar h_{c_j} iv_- \cdot D_{c_j} \frac{\s n_+}{2} h_{c_j}  - \frac{1}{4} F_{c_j \mu \nu}^A F_{c_j}^{\mu \nu A} + (c_j \leftrightarrow \bar c_j) + \f L_{us_j}.
\end{align}
As before, the $c_{j+1}$ modes can interact with the $\bar c_j$ modes through ``messenger'' interactions generated by
\begin{align}
i v_+ \cdot D_{\bar c_j}(x) \rightarrow i v_+ \cdot D_{\bar c_j}(x) + i g\, \frac{n_- \cdot v_+}{2} n_+ \cdot A_{c_{j+1}}(x_+),
\end{align}
which must be removed by performing a decoupling transformation 
\begin{align} \label{eq: j decoupling}
&h_{c_j}(x) \rightarrow W_{\bar c_{j+1}}(x_-) h_{c_j}(x), \qquad A_{c_{j+1}}^{\mu}(x) \rightarrow W_{\bar c_{j+1}}(x_-) A_{c_{j+1}}^{\mu}(x) W_{\bar c_{j+1}}^{\dagger}(x_-),
\\
&h_{\bar c_j}(x) \rightarrow W_{c_{j+1}}(x_+) h_{\bar c_j}(x), \qquad A_{\bar c_{j+1}}^{\mu}(x) \rightarrow W_{c_{j+1}}(x_+) A_{\bar c_{j+1}}^{\mu}(x) W_{c_{j+1}}^{\dagger}(x_+).
\end{align}
We can now proceed inductively to obtain the desired infinite series of EFTs that are summarized in the table \ref{tab: 1}.
\begin{table}
\begin{center}
\begin{tabular}{ |c|c|c|c|c|c|c| } 
 \hline
 Theory & QCD & SCET$_{\rm II}$ & bHQET$_1$ &  ... & bHQET$_j$ & ...  \\ 
 Scale & $Q^2$ & $m^2$ & $m^4/Q^2$ & ... & $m^{2j+2}/Q^{2j}$ & ... \\ 
 Fields & $\psi, A$ & $\xi_c, A_c, \xi_{\bar c}, A_{\bar c}, q_s, A_s$ & $h_{c_1}, A_{c_1}, h_{\bar c_1}, A_{\bar c_1} $ & ... & $h_{c_j}, A_{c_j}, h_{\bar c_j}, A_{\bar c_j} $ & ... \\ 
 \hline
\end{tabular}
\end{center}
\caption{Sequence of EFTs. Note that the intermediate SCET$_{\rm I}$ theory at the scale $\mu^2\sim Qm$ has not been shown in this table. We have also not included the $A_{c_{j+1}}, A_{\bar c_{j+1}}$ within bHQET$_j$, since we have assumed that the decoupling transformation in eq. \eqref{eq: j decoupling} has been performed.}
\label{tab: 1}
\end{table}

Additionally, as remarked before, there also exist an infinite tower of ultra-soft modes with scalings
\begin{equation}
p_{us_j} \sim (\lambda,\lambda,\lambda)\,\lambda^{2j}\,, \qquad j \in \mathbb{N}\,,
\end{equation}
corresponding to successive steps in the EFT cascade below the soft scale. The standard ultra-soft mode corresponds only to the top of this tower. In contrast, the modes relevant for our analysis in later sections are those at the first non-trivial level of this hierarchy, $j=1$, which become physical (i.e.\ non-scaleless) once a regulator introduces an additional infrared scale. This distinction is essential: in pure dimensional regularization the entire tower is scaleless and collapses, whereas with a physical regulator different levels in the tower are resolved and must be treated separately.

The matching coefficient $\mathfrak Z_j$ between bHQET$_j$ and bHQET$_{j+1}$ is defined by
\begin{align}
\bar{h}_{\bar c_{j}} W_{\bar c_{j}} \gamma_\perp^\mu  W_{c_j}^{\dagger} h_{c_j} \equiv 
\mathfrak Z_j\,\,\bar{h}_{\bar c_{j+1}} W_{\bar c_{j+1}} \gamma_\perp^\mu   W_{c_{j+1}}^{\dagger} h_{c_{j+1}}\, 
\end{align}
and the resulting factorization formula is
\begin{align}\label{eq: fact 2}
F_1(Q^2,m^2)
= C(Q^2)\, Z_c^{1/2}(m^2)\, Z_{\bar c}^{1/2}(m^2)\, S(m^2)\,
\prod_{j=1}^{\infty} \mathfrak Z_j
+ \mathcal O(\lambda^2)\,.
\end{align}
In dimensional regularization, the bare matching corrections contributing to $\mathfrak Z_j$ are scaleless, 
\begin{align}
\mathfrak Z_j = 1\,, \qquad j\ge 1\,.
\end{align}
Therefore, the ultra-collinear tower contributes trivially in pure dimensional regularization. As before, the ultra$^j$-soft functions are scaleless and do not contribute either, so we have not shown them explicitly. Hence, \eqref{eq:fact1} follows for the bare form factor and we have shown that even though the ultra$^j$-collinear regions generate non-vanishing contributions diagram by diagram, their net effect cancels to all orders.

The factorization formula~\eqref{eq:fact1} should be understood at the level of \emph{bare} operators. The associated SCET$_{\rm II}$ matrix elements carry infrared singularities that must be separated consistently in the factorized description. The fact that the bare coefficients $\mathfrak Z_j^{\rm bare}=1$ for $j\ge1$ does \emph{not} imply an absence of infrared sensitivity. Rather, dimensional regularization does not resolve scaleless regions and therefore obscures the separation of UV and IR poles. If instead one works with renormalized currents,
$O_j^{\rm bare}=Z_{O_j}(\mu)\,O_j(\mu)$, then the renormalized matching coefficient is
$\mathfrak Z_j(\mu)=\big[Z_{O_{j+1}}(\mu)/Z_{O_j}(\mu)\big]\mathfrak Z_j^{\rm bare}$. Thus, even when $\mathfrak Z_j^{\rm bare}=1$, the matching can acquire a nontrivial pole structure through operator renormalization. 
From the EFT viewpoint, these are UV poles of the low-energy theories, while from the full-theory viewpoint, they reproduce the IR poles of the on-shell form factor, consistent with the general correspondence between IR singularities of on-shell amplitudes and UV renormalization of SCET operators. 
In the abelian limit, this infrared factor exponentiates in a particularly simple way \cite{Yennie:1961ad}, whereas in non-abelian gauge theories exponentiation persists but involves genuinely non-abelian web structures and color correlations. We return to this point in section~\ref{sec: gluon mass reg} where we introduce an explicit infrared regulator (massive gauge boson), which endows the ultra-collinear region with a scale and makes the UV/IR bookkeeping manifest.

More broadly, the EFT tower is guided by the region analysis of the underlying loop integrals. While a general all-order prescription for identifying the complete set of relevant Minkowski regions is still under development; see Ref.~\cite{Ma:2026pjx} for recent progress, our conclusions do not depend on such a classification. Instead, we show that ultra-collinear contributions, once isolated, do not correspond to independent dynamical modes but are already encoded in the soft–collinear structure enforced by gauge invariance.

\subsection{Illustrative graphical examples}
\label{sec:examples}

In this section, we provide some explicit multiloop examples showcasing the factorization of ultra-collinear physics.
\begin{figure}
    \centering
    \includegraphics[width=1\linewidth]{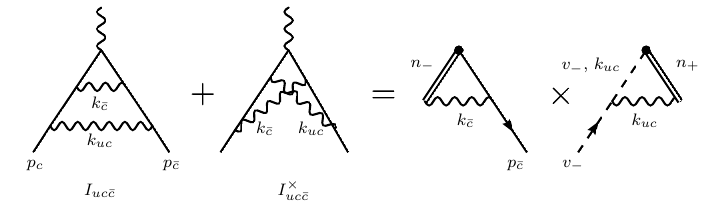}
    \caption{Example of Ward identity cancellation at two loops. The two leftmost diagrams denote specific regions of QCD diagrams. The second diagram from the right is a contribution to $\sqrt{Z}$ in SCET$_{\rm II}$. The rightmost diagram is a contribution to $\sqrt{\mathfrak Z_1}$ in bHQET$_1$. Heavy fermion lines in bHQET$_1$ are denoted by dashed lines. Wilson lines are denoted by double lines.}
    \label{fig: 2loop uc 1}
\end{figure}
We start by considering the regions of the two leftmost graphs involving ultra-collinear momenta shown in fig.~\ref{fig: 2loop uc 1}:
\begin{align}
I_{uc \bar c} &= \int  \frac{ d^dk_{\bar c} d^dk_{uc} \, n_-p_{\bar c} \,  n_- (p_{\bar c} + k_{\bar c}) \, n_+ p_c \, n_+ p_c }{k_{\bar c}^2 k_{uc}^2 [n_- p_{\bar c} \, n_+ k_{uc}] [k_{\bar c}^2 + 2 p_{\bar c} k_{\bar c} +  n_- (p_{\bar c} + k_{\bar c}) \, n_+k_{uc} ] [n_+ p_c \, n_- k_{\bar c}] [2p_c k_{uc}]  },
\\
I_{uc \bar c}^{\times} &= \int \frac{d^dk_{\bar c} d^dk_{uc} \, n_-(p_{\bar c} + k_{\bar c}) \, n_-(p_{\bar c} + k_{\bar c}) \, n_+ p_c \, n_+ p_c}{k_{\bar c}^2 k_{uc}^2 [k_{\bar c}^2 + 2p_{\bar c} k_{\bar c}] [k_{\bar c}^2 + 2 p_{\bar c} k_{\bar c} +  n_- (p_{\bar c} + k_{\bar c}) \, n_+k_{uc} ] [n_+p_c\, n_- k_{\bar c}] [2p_c k_{uc}]},
\end{align}
where an overall factor of $\frac{4 g_0^4}{(2\pi)^{2d}} \bar u_{\bar c} \gamma_{\perp}^{\mu} u_c  $ has been omitted for brevity.
By explicit calculation, it was found in \cite{terHoeve:2023ehm} that
\begin{align}
I_{uc \bar c} = - I_{uc \bar c}^{\times}.
\end{align}
The result for $I_{uc \bar c}$ is given in eq. (5.22) of \cite{terHoeve:2023ehm} and the result for $I_{uc \bar c}^{\times}$ is given in eq. (5.50). This cancellation can be readily observed at the integrand level. Adding the two diagrams, we obtain:
\begin{align}
\frac{n_- p_{\bar c}}{ n_- p_{\bar c} \, n_+ k_{uc}} + \frac{n_-(p_{\bar c} + k_{\bar c})}{k_{\bar c}^2 + 2 p_{\bar c} k_{\bar c}} = \frac{k_{\bar c}^2 + 2p_{\bar c} k_{\bar c} + n_-(p_{\bar c} + k_{\bar c})(n_+ k_{uc})}{n_+ k_{uc} (k_{\bar c}^2 + 2p_{\bar c} k_{\bar c})}.
\end{align}
The numerator cancels with the denominator of the $p_{\bar c} + k_{\bar c} + k_{uc}$ propagator, resulting in a scaleless integral:
\begin{align}
I_{uc \bar c} + I_{uc\bar c}^{\times} = 
\int \frac{ d^dk_{\bar c} }{(2\pi)^d} \frac{ d^dk_{uc} }{(2\pi)^d} \, \frac{   n_- (p_{\bar c} + k_{\bar c}) \, n_+ p_c \, n_+ p_c }{k_{\bar c}^2 k_{uc}^2 [ n_+ k_{uc} ] [k_{\bar c}^2 + 2 p_{\bar c} k_{\bar c}] [n_+ p_c \, n_- k_{\bar c}] [2p_c k_{uc}]  } = 0.
\end{align}
The cancellation is not accidental but a consequence of the Ward identity. It occurs after summing the two possible insertions of the ultra-collinear gluon into the anti-collinear fermion line. 

\begin{figure}
    \centering
    \includegraphics[width=0.9\linewidth]{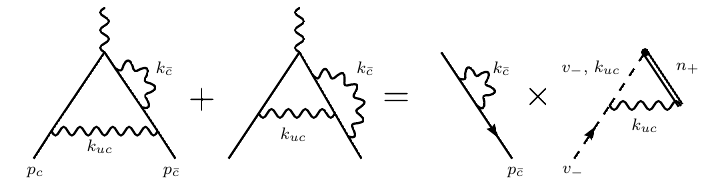}
    \caption{Another two-loop example of Ward-identity cancellation. The two leftmost diagrams
denote the ultra-collinear region of the corresponding QCD graphs. The second diagram
from the right is the SCET$_{\rm II}$ contribution to $\sqrt{Z}$, and the rightmost diagram is the
bHQET$_1$ contribution to $\sqrt{\mathfrak Z_1}$. Heavy-fermion lines in bHQET$_1$ are dashed and Wilson
lines are drawn as double lines.}
    \label{fig: 2loop uc 2}
\end{figure}

\begin{figure}
    \centering
    \includegraphics[width=1\linewidth]{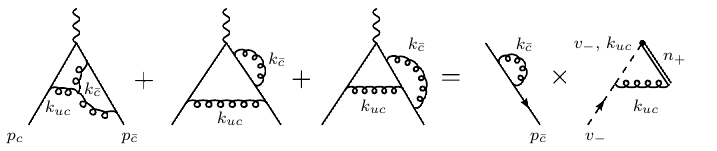}
    \caption{Two-loop example of Ward-identity cancellation in a non-abelian gauge theory. The three
leftmost diagrams denote the ultra-collinear regions of the corresponding QCD graphs. The
second diagram from the right is the SCET$_{\rm II}$ contribution to $\sqrt{Z}$, and the rightmost
diagram is the bHQET$_1$ contribution to $\sqrt{\mathfrak Z_1}$.}
    \label{fig: 2loop uc 3}
\end{figure}

The Ward identity is built into the EFT formalism by maintaining manifest gauge invariance of the effective Lagrangians. The factor
\begin{align}
\int d^dk_{\bar c} \frac{n_-(p_{\bar c} + k_{\bar c}) }{k_{\bar c}^2 [k_{\bar c}^2 + 2p_{\bar c} k_{\bar c}] [n_- k_{\bar c}]}
\end{align}
is part of the (anti-)collinear function $\sqrt{Z_c}$, which is integrated out from the viewpoint of the lower energy theory bHQET$_1$. The remaining (scaleless) integral is
\begin{align}
\int d^dk_{uc} \frac{1}{k_{uc}^2 \, v_-k_{uc} \, n_+ k_{uc} } \subset \bra {0} W_{uc}^{\dagger} h_{uc} \ket{p_c}.
\end{align}
A second example is shown in fig.~\ref{fig: 2loop uc 2}. The individual results for the two leftmost diagrams in this figure can be found in \cite{terHoeve:2023ehm}, eqs. (5.14) and (5.41), and a similar cancellation follows. These arguments extend analogously to non-abelian cases, and we give an example in fig.~\ref{fig: 2loop uc 3}. Following \cite{terHoeve:2023ehm}, the second diagram carries a color factor $C_F^2$, while the third diagram carries a color factor $C_F^2-\frac12C_F C_A$. The cancellation of the ``abelian'' part (i.e. the $C_F^2$ part) proceeds exactly as in the QED case, while the ``non-abelian'' $C_F C_A$ contribution cancels similarly to the three-gluon-vertex contribution from the first diagram. As before, the factorization into a collinear and a (scaleless) ultra-collinear integral can be arranged at the integrand level after summing over the three diagrams.
\begin{figure}
    \centering
    \includegraphics[width=1\linewidth]{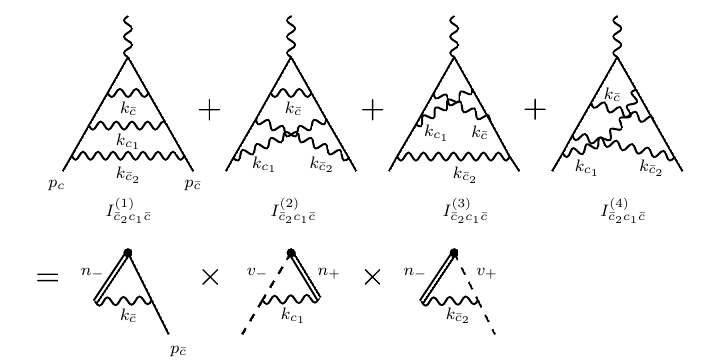}
\caption{Three-loop example of Ward-identity cancellation. The upper row shows the relevant
three-loop configurations/regions involving ultra-collinear momenta. The lower row shows the EFT
representation of the factorized structure: the left diagram gives the SCET$_{\rm II}$ contribution to
$\sqrt{Z}$, the middle diagram the bHQET$_1$ contribution to $\sqrt{\mathfrak Z_1}$, and the right diagram the
bHQET$_2$ contribution to $\sqrt{\mathfrak Z_2}$. }
    \label{fig: 3loop uc}
\end{figure}

Finally, we discuss a three-loop example, see fig. \ref{fig: 3loop uc}.
Consider the two leftmost graphs in the upper row of fig. \ref{fig: 3loop uc}
\begin{align} \notag
I_{\bar c_2 c_1 \bar c}^{(1)} &= \int \frac{ d^dk_{\bar c} d^dk_{c_1} d^dk_{\bar c_2} }{k_{\bar c}^2 k_{c_1}^2 k_{\bar c_2}^2 }    \frac{n_- p_{\bar c} \, n_- p_{\bar c} \, n_- (p_{\bar c} + k_{\bar c})}{[2p_{\bar c}k_{\bar c_2}] [n_- p_{\bar c} \, n_+ k_{c_1}] [k_{\bar c}^2 + 2 p_{\bar c} k_{\bar c} + n_-(p_{\bar c} + k_{\bar c} ) n_+ k_{c_1}]},
\\ 
&\quad \times  \frac{ n_+ p_c \, n_+ p_c \, n_+ p_c }{[n_+ p_c \, n_- k_{\bar c}] [2p_c k_{c_1} + n_+ p_c \, n_- k_{\bar c_2}] [n_+ p_c \, n_- k_{\bar c_2}]}
\\ \notag
I_{\bar c_2 c_1 \bar c}^{(2)} &= \int \frac{ d^dk_{\bar c} d^dk_{c_1} d^dk_{\bar c_2} }{k_{\bar c}^2 k_{c_1}^2 k_{\bar c_2}^2 }    \frac{n_- p_{\bar c} \, n_- p_{\bar c} \, n_- (p_{\bar c} + k_{\bar c})}{[2p_{\bar c}k_{\bar c_2}] [n_- p_{\bar c} \, n_+ k_{c_1}] [k_{\bar c}^2 + 2 p_{\bar c} k_{\bar c} + n_-(p_{\bar c} + k_{\bar c} ) n_+ k_{c_1}]}
\\ 
&\quad \times  \frac{ n_+ p_c \, n_+ p_c \, n_+ p_c }{[n_+ p_c \, n_- k_{\bar c}] [2p_c k_{c_1} + n_+ p_c \, n_- k_{\bar c_2}] [2p_c k_{c_1}]},
\end{align}
where we have neglected the overall factor $\frac{8g_0^4}{(2\pi)^{3d}} \bar u_{\bar c} \gamma_{\perp}^{\mu} u_c$. One readily finds that both integrals are scaleful. 
Adding those together cancels the denominator of the $p_c + k_{c_1} + k_{\bar c_2}$ propagator. 
\begin{align} \notag
I_{\bar c_2 c_1 \bar c}^{(1)} + I_{\bar c_2 c_1 \bar c}^{(2)} &= \int \frac{ d^dk_{\bar c} d^dk_{c_1} d^dk_{\bar c_2} }{k_{\bar c}^2 k_{c_1}^2 k_{\bar c_2}^2 }   \frac{n_+ p_c  }{[n_- k_{\bar c_2}] [2p_c k_{c_1}]}
\\
&\quad \times \frac{n_- p_{\bar c} \, n_- p_{\bar c} \, n_- (p_{\bar c} + k_{\bar c})}{ [n_- k_{\bar c}][2p_{\bar c}k_{\bar c_2}] [n_- p_{\bar c} \, n_+ k_{c_1}] [k_{\bar c}^2 + 2 p_{\bar c} k_{\bar c} + n_-(p_{\bar c} + k_{\bar c} ) n_+ k_{c_1}]}. 
\end{align}
Note that the $k_{\bar c_2}$ integral is already scaleless after adding these two graphs, so that $I_{\bar c_2 c_1 \bar c}^{(1)} + I_{\bar c_2 c_1 \bar c}^{(2)}  = 0$. However, we have not yet obtained the fully factorized form, which requires adding the two rightmost graphs in the upper row of fig.~\ref{fig: 3loop uc}. Those read
\begin{align} \notag
I_{\bar c_2 c_1 \bar c}^{(3)} &= \int \frac{ d^dk_{\bar c} d^dk_{c_1} d^dk_{\bar c_2} }{k_{\bar c}^2 k_{c_1}^2 k_{\bar c_2}^2 }    \frac{n_- p_{\bar c} \, n_- (p_{\bar c} + k_{\bar c}) \, n_- (p_{\bar c} + k_{\bar c})}{[2p_{\bar c}k_{\bar c_2}] [k_{\bar c}^2 + 2 p_{\bar c} k_{\bar c}] [k_{\bar c}^2 + 2 p_{\bar c} k_{\bar c} + n_-(p_{\bar c} + k_{\bar c} ) n_+ k_{c_1}]}
\\
&\quad \times  \frac{ n_+ p_c \, n_+ p_c \, n_+ p_c }{[n_+ p_c \, n_- k_{\bar c}] [2p_c k_{c_1} + n_+ p_c \, n_- k_{\bar c_2}] [n_+ p_c \, n_- k_{\bar c_2}]},
\\ \notag
I_{\bar c_2 c_1 \bar c}^{(4)} &= \int \frac{ d^dk_{\bar c} d^dk_{c_1} d^dk_{\bar c_2} }{k_{\bar c}^2 k_{c_1}^2 k_{\bar c_2}^2 }    \frac{n_- p_{\bar c} \, n_- (p_{\bar c} + k_{\bar c}) \, n_- (p_{\bar c} + k_{\bar c})}{[2p_{\bar c}k_{\bar c_2}] [k_{\bar c}^2 + 2 p_{\bar c} k_{\bar c}] [k_{\bar c}^2 + 2 p_{\bar c} k_{\bar c} + n_-(p_{\bar c} + k_{\bar c} ) n_+ k_{c_1}]}
\\
&\quad \times  \frac{ n_+ p_c \, n_+ p_c \, n_+ p_c }{[n_+ p_c \, n_- k_{\bar c}] [2p_c k_{c_1} + n_+ p_c \, n_- k_{\bar c_2}] [2p_c k_{c_1}]},
\end{align}
and add up to
\begin{align} \notag
I_{\bar c_2 c_1 \bar c}^{(3)} + I_{\bar c_2 c_1 \bar c}^{(4)} &= \int \frac{ d^dk_{\bar c} d^dk_{c_1} d^dk_{\bar c_2} }{k_{\bar c}^2 k_{c_1}^2 k_{\bar c_2}^2 }   \frac{ n_+ p_c  }{[n_- k_{\bar c_2}] [2p_c k_{c_1}]}
\\
&\quad \times \frac{n_- p_{\bar c} \, n_- (p_{\bar c} + k_{\bar c}) \, n_- (p_{\bar c} + k_{\bar c})}{[n_- k_{\bar c}][2p_{\bar c}k_{\bar c_2}] [k_{\bar c}^2 + 2 p_{\bar c} k_{\bar c}] [k_{\bar c}^2 + 2 p_{\bar c} k_{\bar c} + n_-(p_{\bar c} + k_{\bar c} ) n_+ k_{c_1}]} . 
\end{align}
Thus, the sum has the desired factorized form
\begin{align} \notag
&I_{\bar c_2 c_1 \bar c}^{(1)} + I_{\bar c_2 c_1 \bar c}^{(2)} + I_{\bar c_2 c_1 \bar c}^{(3)} + I_{\bar c_2 c_1 \bar c}^{(4)}
\\
&= \underbrace{  \int d^d k_{\bar c} \frac{n_- (p_{\bar c} + k_{\bar c})}{  k_{\bar c}^2 [k_{\bar c}^2 + 2 p_{\bar c} k_{\bar c}] [ n_- k_{\bar c}]} }_{\subset \bra{p_{\bar c}} \bar \xi_{\bar c} W_{\bar c} \ket 0}\underbrace{  \int d^dk_{c_1} \frac{1}{k_{c_1}^2 [v_- k_{c_1}] [n_+ k_{c_1}] } }_{\subset \bra 0 Y_{c_1}^{\dagger} h_{c_1} \ket {p_c} }\underbrace{ \int  d^dk_{\bar c_2} \frac{ 1}{k_{\bar c_2}^2 [v_+ k_{\bar c_2}][n_- k_{\bar c_2}] } }_{\subset \bra{ p_{\bar c} } \bar h_{ \bar c_2} Y_{\bar c_2} \ket 0 }. 
\end{align}
This completes our demonstration of the Ward-identity cancellation at three loops.

\section{Explicit computations of the soft and jet functions}
\label{sec:rapreg}

\subsection{Definition}

The soft and collinear functions in eq.~\eqref{eq:fact1} contain rapidity divergences and are therefore ill-defined without additional regularization. We adopt the $\eta$ rapidity regulator of Ref.~\cite{Chiu:2012ir}, implemented by modifying the Wilson lines in eqs.~\eqref{eq: soft wline 1}, \eqref{eq: coll wline 1}, and \eqref{eq: acoll wline 1} as
\begin{align} \notag
S_{\pm}(x) &= \mathcal{ P} \exp\left ( i g_0\, w(\nu)\, \nu^{\eta/2} \int_{-\infty}^0 ds\,  n_{\pm \mu} \bigl(|2 i \p^z|^{-\eta/2}A_s^{\mu}\bigr)(x+sn_{\pm}) \right ),
\\ \label{eq: eta regulator}
W_c(x) &= \mathcal{ P}  \exp\left ( i g_0\, w^2(\nu)\, \nu^{\eta} \int_{-\infty}^0 ds\, n_{+\mu} \bigl(|in_+ \p|^{-\eta}A_{c}^{\mu}\bigr)(x+sn_+) \right ),
\\ \notag
W_{\bar c}(x) &= \mathcal{ P} \exp\left ( i g_0\, w^2(\nu)\, \nu^{\eta} \int_{-\infty}^0 ds\, n_{-\mu} \bigl(|in_- \p|^{-\eta}A_{\bar c}^{\mu}\bigr)(x+sn_-) \right ).
\end{align}
Here, $\eta$ acts as the regulator, $\nu$ is the associated rapidity scale, and $w(\nu)$ is a bookkeeping
parameter used to derive rapidity RG equations. In momentum space,  derivative insertions
$|2 i\p^z|^{-\eta/2}$ and $|i n_\pm \p|^{-\eta}$ generate sector-dependent weights (e.g. $|2k^z|^{-\eta/2}$ in the
soft function) that regulate rapidity divergences, which appear as poles in $1/\eta$.  Our prescription is to choose $w(\nu) = \nu^{-\eta/2}$, and we expand in $\eta$ first (i.e. take $\eta\to 0$ before $\epsilon\to0$). This ordering is essential for consistent extraction of rapidity divergences. The asymmetric assignment: $w(\nu)\nu^{\eta/2}$ per soft Wilson line versus $w^2(\nu)\nu^\eta$ for each collinear Wilson line is chosen so that the rapidity anomalous dimensions $\gamma_S^{(\nu)}$ and $\gamma_Z^{(\nu)}$ are equal and opposite, guaranteeing that the product of collinear and soft functions is $\nu$-independent to all orders~\cite{Chiu:2012ir}.
We work in the center-of-mass frame with light-cone vectors $n_\pm=(1,0,0,\pm1)$.
In this basis,  $\partial^z=\tfrac12(n_-\partial-n_+\partial)$, so the soft weight $|2i\partial^z|^{-\eta/2}$
coincides with the standard $|in_+\partial-i n_-\partial|^{-\eta/2}$ form of the $\eta$ regulator.

We compute the soft factor $S$, defined in \eqref{eq:Sdef}, through the first non-vanishing order. In pure dimensional regularization, the one-loop contribution is scaleless and vanishes, so the leading correction arises at two loops. Moreover, all two-loop graphs containing only massless gluons (and potentially massless fermions) are scaleless. The only nonzero contribution comes from diagrams with  massive fermion bubble insertions. The two-loop soft factor is known~\cite{Hoang:2015vua}, but the derivation we present here follows a different approach, using the $\eta$ regulator rather than the analytic regulator, which facilitates clean NNLL resummation. Together with NNLO results for $F_1$ and the hard matching coefficient $C(Q^2)$, we will use $S$ computed below to extract the combined collinear-anti-collinear contribution
\begin{align}\label{eq:Zdef}
Z(m^2,\mu,\nu) \equiv Z^{1/2}_c(m^2,\mu,\nu)\, Z^{1/2}_{\bar c}(m^2,\mu,\nu)
\end{align}
at NNLO. The manifest symmetry of the $\eta$-regulator ensures that $Z_c = Z_{\bar c}$. 

Finally, we comment on the overlap subtraction associated with the soft--collinear boundary in SCET$_{\rm II}$ (often called the soft-bin or zero-bin) \cite{Manohar:2006nz,Idilbi:2007ff}. In SCET, overlap contributions are removed by expanding the integrand in the scaling of the overlapping mode and subtracting the resulting contribution integrated over all momenta. In the $\eta$-regulator scheme used here, the rapidity regulator itself is defined as the appropriate mode-dependent limit of the operator insertion. In a collinear limit, the soft factor $|2i \partial^z|^{-\eta}$ reduces to the large light-cone component (schematically $|2i \partial^z|^{-\eta}\to |i n_+\partial|^{-\eta}$). With this prescription, the soft-bin integrals are scaleless and vanish in pure dimensional regularization \cite{Manohar:2006nz}, so no explicit zero-bin subtraction is required for the soft function.

\subsection{Calculation of $S$ and $Z$} \label{sec: soft}
\begin{figure}
    \centering
    \includegraphics[width=\linewidth]{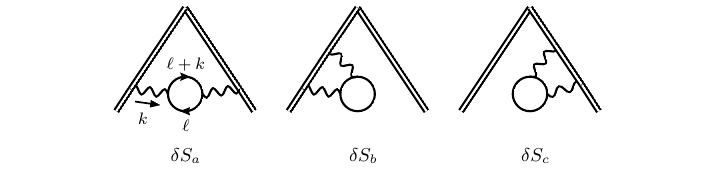}
   \caption{Non-vanishing two-loop graphs contributing to the soft factor $S$ in the
presence of massive fermions. Each diagram contains a one-loop massive-fermion vacuum-polarization
subgraph in the exchanged gauge boson.}
    \label{fig: S}
\end{figure}
We calculate the two-loop contribution to $S = \frac{1}{N_c} \text{tr} \bra 0 S_-^{\dagger} S_+ \ket 0$, which we denote by $S^{(2)}$. The three non-zero graphs are shown in fig.~\ref{fig: S}, and they all have the one-loop quark-loop vacuum polarization $\Pi_{\mu \nu}$ as a subgraph. By the Ward identity
\begin{align} \label{eq: Pi WI}
\Pi_{\mu \nu}(k, m^2) = \Pi(k^2,m^2) \left ( g_{\mu \nu} - \frac{k_{\mu} k_{\nu}}{k^2} \right ),
\end{align}
where the one-loop vacuum polarization reads
\begin{align} \label{eq: Pi 1loop}
\Pi(k^2,m^2) &= - 8 T_F  \frac{\Gamma(\epsilon)}{(4\pi)^{2-\epsilon}} \int_0^1 du\,  u (1-u) \left [m^2 - u (1-u) k^2\right ]^{-\epsilon}.
\end{align}
Here $T_F$ is the fundamental-representation trace normalization of the  gauge group generators, $\text{tr} \, t^A t^B = T_F \delta^{AB}$, and $C_A$ is the adjoint-representation Casimir. The QED case is obtained by setting $T_F = 1, \, C_A = 0, \, C_F = 1$.
The first graph in fig. \ref{fig: S} gives
\begin{align}
 S^{(2)}_1 = - C_F T_F \frac{g_0^4 w^2}{(4\pi)^{d/2}} \nu^{\eta} \int \frac{d^dk}{i\pi^{d/2}} \frac{1}{n_+k} \frac{1}{n_-k } \frac{1}{k^2} n_+^{\mu} n_-^{\nu} \, |2k^z|^{-\eta} \,  \Pi_{\mu \nu}(k, m_0^2),
\end{align}
where we now distinguish the bare mass $m_0$ from the pole mass $m$.
 The longitudinal contribution $\frac{k_{\mu} k_{\nu}}{k^2}$ of $\Pi_{\mu \nu}$ in $S^{(2)}_1$ is canceled by Wilson line self energy graphs $S^{(2)}_2 + S^{(2)}_3$ up to terms of $\f O(\eta)$, as required by gauge invariance. Thus, we obtain
\begin{align} \label{eq: delta S interm 0}
 S^{(2)} &= - C_F T_F  \frac{g_0^4 w^2 \nu^{\eta} }{(4\pi)^{d/2}}\int \frac{d^dk}{i \pi^{d/2}} \frac{2 \, |2k^z|^{-\eta} \,   \Pi(k^2, m_0^2 - i0)}{(n_+ k + i0) \, (n_-k + i0) \, ( k^2 + i0)}. 
\end{align}
where we have restored the $i0$ prescription. Carrying out the $k_\perp$ integration, we obtain
\begin{align} \notag
S^{(2)} &= \frac{\alpha_0^2}{\pi^2} C_F T_F (4\pi)^{2\epsilon} w^2 \nu^{\eta} \Gamma(2\epsilon) \int_0^1 \frac{dv}{v^{1-\epsilon}} \int_0^1 du \, [u (1-u)]^{1-\epsilon} 
\\ \label{eq: delta S interm 1}
&\quad \times\frac{1}{2\pi i}\int \frac{ dn_+ k dn_-k \, |2k^z|^{-\eta}}{(n_+k + i0) \, (n_-k + i0) (- n_+ k \, n_- k + \frac{v m_0^2}{u (1-u) } - i0 )^{2\epsilon} }.
\end{align}
The expression in the lower line of eq. \eqref{eq: delta S interm 1} can be evaluated as
\begin{align} \notag
& \int \frac{dk^0 dk^z \, \frac{1}{i\pi} \, |2k^z|^{-\eta}}{(k^0 + k^z + i0)(k^0 - k^z + i0) \left (- k^0 + \sqrt{(k^z)^2 + \frac{vm_0^2}{u(1-u)}} - i0\right )^{2\epsilon} \left (k^0 + \sqrt{(k^z)^2 + \frac{vm_0^2}{u(1-u)}} - i0\right )^{2\epsilon}}
\\
&\quad = \frac{2^{1-\eta} \Gamma(1/2-\eta/2)  \Gamma(2\epsilon + \eta/2)}{\eta \sqrt{\pi} \Gamma(2\epsilon)}   \left (\frac{vm_0^2}{u(1-u)} \right )^{-2\epsilon - \eta/2},
\end{align}
by first wrapping the $k^0$ contour around the branch cut $k^0 < - \sqrt{(k^z)^2 + \frac{vm_0^2}{u(1-u)}}$, after which the remaining integral can be performed using elementary methods. The $u$ and $v$ integrals can be readily evaluated, and we finally obtain for the \textit{bare} soft function
\begin{align} \notag
S &= 1 +  \left ( \frac{\alpha_0}{4\pi} \right )^2\left ( \frac{4\pi e^{-\gamma_E}}{m^2} \right )^{2\epsilon} w^2(\nu)\left ( \frac{\nu^2}{m^2} \right )^{\eta/2}   C_F T_F \Bigg \{ - \frac{8}{3\eta}  (\Gamma(\epsilon)e^{\epsilon \gamma_E} )^2  \frac{1+\epsilon}{1+\frac{8}{3}\epsilon + \frac{4}{3}\epsilon^2} 
\\ \label{eq: soft result}
&\qquad + \frac{2}{\epsilon^3} - \frac{10}{9\epsilon^2}   + \frac{1}{\epsilon} \left ( - \frac{56}{27} + \frac{\pi^2}{3} \right ) + \frac{328}{27} - \frac{5\pi^2}{27} + 4 \zeta_3 + \f O(\eta, \epsilon) \Bigg \} + \f O(\alpha_0^3).
\end{align}
Eq. \eqref{eq:fact1} allows us to determine $Z$ at NNLO using the known full result of the Sudakov form factor, which we provide in Appendix~\ref{app:F1NNLO}. We obtain the bare collinear function
\begin{align} \notag
Z &= 1 + \frac{\alpha_0}{4\pi} \left ( \frac{4\pi e^{-\gamma_E}}{m^2} \right )^{\epsilon} C_F Z^{(1)}
\\ \label{eq: Z0 twoloop}
&\quad + \left ( \frac{\alpha_0}{4\pi} \right )^2 \left ( \frac{4\pi e^{-\gamma_E}}{m^2} \right )^{2\epsilon} w^2(\nu) \left ( \frac{\nu^2}{Q^2} \right )^{\eta/2} C_F \, \Bigg \{ C_F Z_F^{(2)} + C_A  Z_A^{(2)}  + T_F Z_T^{(2)} + \f O(\eta ; \epsilon) \Bigg \}  
\\ \notag
&\quad + \f O(\alpha_0^3),
\end{align}
where
\begin{align} 
Z^{(1)} &= \frac{2}{\epsilon^2} + \frac{1}{\epsilon} + 4 + \frac{\pi^2}{6} + \epsilon \left ( 8 + \frac{\pi^2}{12} - \frac{2}{3} \zeta_3 \right )  + \epsilon^2 \left ( 16 + \frac{\pi^2}{3} - \frac{1}{3} \zeta_3  + \frac{\pi^4}{80} \right ) + \f O(\epsilon^3),
\\ \notag
Z_F^{(2)} &= \frac{2}{\epsilon^4} + \frac{2}{\epsilon^3} + \frac{1}{\epsilon^2} \left ( \frac{17}{2} + \frac{\pi^2}{3}\right )  + \frac{1}{\epsilon} \left (\frac{83}{4} - \frac{2\pi^2}{3} + \frac{32}{3} \zeta_3  \right ) 
\\
&\qquad + \frac{561}{8} + \frac{61\pi^2}{12} - \frac{22}{3} \zeta_3 - \frac{77\pi^4}{180} - 8\pi^2 \ln 2,
\\ \notag
Z_A^{(2)} &= \frac{11}{6\epsilon^3} + \frac{1}{\epsilon^2} \left ( \frac{50}{9} - \frac{\pi^2}{6} \right ) + \frac{1}{\epsilon} \left ( \frac{1957}{108} + \frac{67 \pi^2}{36} - 15 \zeta_3 \right ) 
\\
&\qquad + \frac{31885}{648} + \frac{89\pi^2}{27} + \frac{67}{9} \zeta_3 - \frac{47  \pi^4}{180} + 4 \pi^2 \ln 2,
\\ \notag
Z_T^{(2)} &= \frac{8}{3\eta}  (\Gamma(\epsilon)e^{\epsilon \gamma_E} )^2  \frac{1+\epsilon}{1+\frac{8}{3}\epsilon + \frac{4}{3}\epsilon^2}  - \frac{8}{3 \epsilon^3} + \frac{2}{3\epsilon^2} + \frac{1}{\epsilon} \left ( - \frac{17}{3} - \frac{8\pi^2}{9} \right ) 
\\
&\qquad + \frac{1411}{162} - \frac{11\pi^2}{9} - \frac{32}{9} \zeta_3.
\end{align}

\subsection{Renormalization and resummation} 
\label{sec: resummation 1}

First, we renormalize the coupling according to the $\overline{{\rm MS}}$ prescription. The bare coupling $\alpha_{0}$ and renormalized coupling are related by
\begin{align}
\alpha_0 = \alpha(\mu) \left ( \frac{\mu^2 e^{\gamma_E}}{4\pi} \right )^{\epsilon}
\Bigg \{ 1 - \frac{\alpha(\mu)}{4\pi}\, e^{\epsilon \gamma_E}\Gamma(\epsilon)\, \beta_0 + \f O(\alpha^2) \Bigg \},
\qquad
\beta_0 = \frac{11}{3}C_A - \frac{4}{3}T_F\, n_f .
\end{align}
Here, $n_f$ denotes the total number of active quark flavors at the scale where the coupling is evaluated.
Expressions for other schemes can be readily obtained using the results in section~\ref{sec: soft} and appendix \ref{app:F1NNLO}. In the remainder of this subsection we set $n_f = 1$.

With pure dimensional regularization, the bare functions $S$ and $Z$ contain overlapping UV/IR poles,
so the separation of ``UV'' versus ``IR'' poles at the level of individual factors is scheme dependent unless an additional infrared regulator is introduced. We therefore
define ${\rm Z}_S$ and ${\rm Z}_Z$ by minimal subtraction of the $1/\epsilon$ poles of the corresponding bare
matrix elements; the resulting anomalous dimensions then retain infrared information, while only the
combined product in eq.~\eqref{eq: Z product} is physical. 

We also absorb the IR divergences in $F_1$ into a multiplicative factor. We define the ``renormalized'' functions $F_1^{\rm fin}, C_q^R, Z_q^R , S_q^R$ as
\begin{align}
F_1 = F_1^{\rm IR} F_1^{\rm fin}, \qquad Z = {\rm Z}_Z Z_q^R, \qquad S = {\rm Z}_S S_q^R, \qquad C = {\rm Z}_C C_q^R,
\end{align}
where the IR divergent part is 
\begin{align} \label{eq: Z product}
{\rm Z}_Z {\rm Z}_S {\rm Z}_C = F_1^{\rm IR}.
\end{align}
${\rm Z}_C$ is the inverse of the standard renormalization factor of the vector current in SCET.\footnote{Our notation slightly departs from the  convention in \cite{Beneke:2017ztn,Beneke:2018rbh}, where the renormalization factor is defined such that it multiplies the operator, whereas here ${\rm Z}_C$ is introduced as a factor multiplying the matching coefficient $C$. As a result, ${\rm Z}_C$ corresponds to the inverse of the usual current renormalization constant. The factors ${\rm Z}_Z$ and ${\rm Z}_S$ follow the standard convention \cite{Beneke:2017ztn,Beneke:2018rbh} of renormalizing operator matrix elements. This choice is purely notational and does not affect any physical results, provided the consistency condition in Eq.~\eqref{eq: Z product} is maintained.}

Since $F_1$ is UV-finite, all UV divergences in the product on the left-hand side of eq. \eqref{eq: Z product} cancel. Furthermore, by specifying the naive $\overline{\rm MS}$ subtraction (of both UV and IR poles) of $C,Z,S$, the scheme subtracting the IR divergences from $F_1$ is determined by eq. \eqref{eq: Z product}. 

For $X\in \{C,Z,S\}$, we define the anomalous dimensions as follows:
\begin{align}
X = {\rm Z}_X X_R, \qquad \gamma_X^{(\mu)} = {\rm Z}_X^{-1} \frac{d}{d\ln \mu} {\rm Z}_X, \qquad \gamma_X^{(\nu)} = {\rm Z}_X^{-1} \frac{d}{d\ln \nu} {\rm Z}_X
\end{align}
and get
\begin{align}  \notag
\gamma_C^{(\mu)} &= \frac{\alpha(\mu)}{\pi} C_F \left ( \ln \frac{\mu^2}{Q^2} + \frac{3}{2} \right ) + \left ( \frac{\alpha(\mu)}{\pi} \right )^2 C_F \Bigg \{ C_F \left [ \frac{3}{16} - \frac{\pi^2}{4} + 3 \zeta_3 \right ] 
\\ \notag
&\quad + C_A \left [ \left ( \frac{67}{36} - \frac{\pi^2}{12} \right ) \ln \frac{\mu^2}{Q^2}  + \frac{961}{432} + \frac{11\pi^2}{144} - \frac{13}{4} \zeta_3  \right ] 
\\ \notag
&\quad + T_F \left ( - \frac{5}{9} \ln \frac{\mu^2}{Q^2} - \frac{65}{108} - \frac{\pi^2}{36} \right ) \Bigg \} + \f O(\alpha^3),
\\ \notag
\gamma_S^{(\mu)} &= \left ( \frac{\alpha(\mu)}{\pi} \right )^2 C_F T_F \Bigg \{\left ( \frac{2}{3} \ln \frac{\mu^2}{m^2} - \frac{5}{9} \right )  \ln \frac{\nu^2}{m^2}  - \ln^2 \frac{\mu^2}{m^2} + \frac{5}{9} \ln \frac{\mu^2}{m^2} + \frac{14}{27} - \frac{\pi^2}{12} \Bigg \} + \f O(\alpha^3),
\\  \notag
\gamma_Z^{(\mu)} &= \frac{\alpha(\mu)}{\pi} C_F \left ( - \ln \frac{\mu^2}{m^2} - \frac{1}{2} \right ) + \left ( \frac{\alpha(\mu)}{\pi} \right )^2 C_F \Bigg \{  C_F \left [ - \frac{3}{16} + \frac{\pi^2}{4} - 3 \zeta_3 \right ]
\\  \label{eq: ADs 1}
&\quad + C_A \left [ \left ( - \frac{67}{36} + \frac{\pi^2}{12} \right ) \ln \frac{\mu^2}{m^2} - \frac{373}{432} - \frac{23\pi^2}{144} + \frac{15}{4} \zeta_3 \right ] 
\\ \notag
&\quad + T_F \left [ \left ( - \frac{2}{3} \ln \frac{\mu^2}{m^2} + \frac{5}{9} \right ) \ln \frac{\nu^2}{Q^2} + \ln^2 \frac{\mu^2}{m^2} - \frac{2}{3} \ln \frac{\mu^2}{m^2} + \frac{1}{12} + \frac{\pi^2}{9} \right ] \Bigg \} + \f O(\alpha^3),
\\ \notag
\gamma_S^{(\nu)} &= -\gamma_Z^{(\nu)} =\left ( \frac{\alpha(\mu)}{\pi} \right )^2 C_F T_F  \Bigg \{ \frac{1}{3} \ln^2 \frac{\mu^2}{m^2} - \frac{5}{9} \ln \frac{\mu^2}{m^2} + \frac{14}{27} + \frac{\pi^2}{36} \Bigg \} + \f O(\alpha^3).
\end{align}
The solution of the evolution equations is straightforward, and we postpone a more detailed discussion until section~\ref{sec:resum}.

\subsection{Additional massive fermions} \label{sec: more quarks}

We now discuss the case of multiple fermion masses in a purely
perturbative setting. We assume a strict hierarchy
\begin{align}
  Q \gg M_1 \gg \cdots \gg M_{n_h} \gg m \gg m_1 \gg \cdots \gg m_{n_\ell} \gg 0\,,
\end{align}
where the external on-shell fermions have mass $m$, the masses $\{M_j\}$ denote additional heavy
fermions that can appear only in virtual loops, and the masses $\{m_i\}$ denote lighter virtual
fermions below the external-mass scale.  This section generalizes the discussion in Sec.~\ref{sec: region cascade general discussion} and makes the resulting tower of effective theories and matching steps explicit.
In particular, lighter fermion masses render some contributions that are scaleless in pure dimensional regularization non-trivial. Each threshold in the hierarchy corresponds to integrating out degrees of freedom, yielding a multiplicative factorization of the form factor into short-distance coefficients and universal low-energy functions.

Lowering the renormalization scale from $\mu \sim Q$, the theory is first
matched from QCD with $n_f=n_h+1+n_\ell$ active fermions onto SCET.  For scales $\mu$ above the external
mass $m$, the leading-power SCET Lagrangian for the collinear modes
associated with the external fermions does not contain a mass term: the
fermion mass is power suppressed and does not enter the dynamics at this
order.  As the scale is lowered, each heavy threshold $\mu\sim M_j$ is crossed
sequentially, and the corresponding fermion is integrated out.  These steps
implement flavor decoupling within SCET; the operator basis remains
unchanged, while the gauge coupling and Wilson coefficients receive matching
corrections.

At the scale $\mu\sim m$, the situation changes qualitatively.  The mass of the external fermion becomes a dynamical scale in the collinear sector, and the $\rm SCET_{II}$ current is matched onto the corresponding bHQET current defined in section~\ref{sec:EFT}.  This matching introduces bHQET Wilson coefficients  now embedded in the full hierarchy of thresholds.  Below $\mu\sim m$, the appropriate description is given by bHQET with $n_\ell$ active fermions.  In the single-flavor case of section~\ref{sec:EFT}, the bHQET matrix elements were scaleless. In the present setup, however, the lighter fermions with masses $m_i$ are integrated out sequentially at their respective thresholds, inducing nontrivial corrections to the bHQET matrix element starting at two loops.

From the perspective of the ultra-collinear tower discussed in section~\ref{sec:EFT}, the presence of lighter fermion masses does not introduce any new dynamical ``interactions'' between modes, but rather promotes previously scaleless contributions into non-trivial matching coefficients. 

With this tower in mind, the factorization structure can be organized
according to the degrees of freedom that are active in each regime. We work with bare, IR divergent objects and do not explicitly write the tower of modes leading to scaleless contributions. 
Consequently, the form factor can be factorized at leading power as
\begin{align}
\notag
F_1(Q^2,\{M_i\},m,\{m_i\}) &=
C^{(1+n_h+n_\ell)}(Q^2)
\left[\prod_{j=1}^{n_h}
S^{(1+ n_h + n_\ell - j)}(M_j^2)\,
\widetilde Z^{(1+ n_h + n_\ell - j)}(M_j^2)\right]
\;
\\ &\quad \times 
S^{(n_\ell)}(m^2)\,
Z^{(n_\ell)}(m^2)
\;
\left[
\prod_{i=1}^{n_\ell}
S^{(n_\ell-i)}(m^2_i) \, \widetilde{ \mathfrak Z }^{(n_\ell - i)}(m^2_i)
\right] 
\label{eq:master-multiscale}
\end{align}
where the product over $M_j$ accounts for the heavy thresholds above the external mass scale, and the product over $m_i$ denotes the successive factors below $m$. The superscripts indicate, in addition to the explicit dependence on the number of massless quarks, the implicit effective  coupling constant. This means that in the object with superscript $^{(n)}$ we replace
\begin{align} 
\label{eq:coupling_replace}
\alpha_0 \;\to\; \alpha_0^{(n)} \equiv \alpha_0 \Bigg[ 1
- \frac{\alpha_0 T_F}{3\pi \epsilon}\,(4\pi)^{\epsilon}\Gamma(1+\epsilon)
\sum_{j=n+1}^{1+n_h+n_\ell} (\mathfrak m_j^2)^{-\epsilon}
+ \mathcal{O}(\alpha_0^2)\Bigg]\,,
\end{align}
where $\{ \mathfrak m_i \} = \{ M_i \} \cup \{ m \} \cup \{m_i \}$. The renormalization can be performed analogously to section~\ref{sec: resummation 1}.

For each intermediate threshold $\mu\sim M_j$, the external mass $m$ is power suppressed, and  the relevant modes are
\begin{align}
k_{s,j} &\sim (M_j,M_j,M_j)\,,\\
k_{c,j} &\sim \Big(Q,\,M_j,\,\frac{M_j^2}{Q}\Big)\,,\qquad
k_{\bar c,j} \sim \Big(\frac{M_j^2}{Q},\,M_j,\,Q\Big)\,,
\end{align}
so that $k_{s,j}^2\sim k_{c,j}^2\sim k_{\bar c,j}^2\sim M_j^2$.
The soft function $S(M_j^2)$ is the SCET soft matrix element associated with soft exchange between
the two collinear sectors, evaluated with soft momenta $k_{s,j}$, while $\widetilde Z(M_j^2)$ denotes
the product of (anti-)collinear factors with loop momenta $k_{c,j},k_{\bar c,j}$ and with the external
legs taken massless and on shell at leading power.

At the final SCET step, $\mu\sim m$, the external mass becomes dynamical in the collinear sectors.
The SCET soft and (anti-)collinear functions are therefore evaluated with the usual massive
$\mathrm{SCET}_{\rm II}$ scalings
\begin{align}
k_s &\sim (m,m,m)\,,\qquad
k_c \sim \Big(Q,\,m,\,\frac{m^2}{Q}\Big)\,,\qquad
k_{\bar c} \sim \Big(\frac{m^2}{Q},\,m,\,Q\Big)\,,
\end{align}
which define $S(m^2)$ and $Z(m^2)$, as in \eqref{eq:Sdef} and \eqref{eq:Zdef}. 

In bHQET, ultra-collinear and ultra-soft modes are dynamical degrees of freedom at a scale $\mu\sim m_i\ll m$. A convenient bookkeeping is to view the ultra-collinear scaling as a rescaled version of the massive collinear scaling,
\begin{align}\label{eq:kusc}
\notag k_{uc,i} &\sim \frac{m_i}{m}\,k_c
\sim \Big(\frac{m_i Q}{m},\,m_i,\,\frac{m_i m}{Q}\Big)\,,\qquad
k_{u\bar c,i} \sim \frac{m_i}{m}\,k_{\bar c}
\sim \Big(\frac{m_i m}{Q},\,m_i,\,\frac{m_i Q}{m}\Big)\,,\\
k_{us,i} &\sim (m_i,m_i,m_i)\,,
\end{align}
so that $k_{uc,i}^2\sim k_{u\bar c,i}^2\sim k_{us,i}^2\sim m_i^2$.
These are the modes whose matrix elements build the bHQET ultra-collinear factors $\w{\mathfrak Z}_i(m_i)$ defined in analogy to \eqref{eq:ZUCdef}, but now endowed with a scale given by light quark masses $m_i$, and the ultra-soft function 
$S(m_i)$ defined as a vacuum matrix element of ultra-soft Wilson lines:
\begin{align} 
 S(m^2_i) = \frac{1}{N_c} \text{tr} \bra 0  S_{-,i}^{\dagger}(0)  S_{+,i}(0) \ket 0,
\end{align}
where
\begin{align}
S_{\pm i }(x) &= \mathcal{P} \exp \left [ ig \int_{-\infty}^0 ds \, n_{\pm} \cdot A_{us,i}(x + s n_{\pm} ) \right ].
\end{align}
Structurally, it is equal to the soft function~\eqref{eq:Sdef}. Hence, we do not employ a new symbol for it (see also subsequent discussion \eqref{eq: frak S def}).

 In terms of the method of regions, the decoupling \eqref{eq:coupling_replace}  arises from the regions of diagrams contributing to $Z$, where the momentum $l$ that flows \emph{inside} the heavy quark loop scales as $l \sim (Q, M_j^2/Q, M_j)$, while the momentum $k$ that flows \emph{into} the loop scales as $k \sim (Q, m^2/Q, m)$. In this configuration, $k$ must be expanded \textit{inside} the loop (but not outside), such that $k^2=0$. The corresponding vacuum polarization subamplitude $\Pi(k^2, M_j^2) \sim \Pi(0, M_j^2)$ then becomes independent of $k$. It is precisely the contribution to the on-shell gluon ${\rm Z}_3$ due to the heavy quark loops, as shown in eq.~\eqref{eq:coupling_replace}. The discussion above applies only to $Z$ and not to $\w Z$ or $S$, since the corresponding contributions in $\w Z$ and $S$ are scaleless.  
 
Eq.~\eqref{eq:master-multiscale} makes explicit the $\ln M_j$ dependence associated with heavy thresholds, and it is precisely these terms that call for resummation via the corresponding factorized functions. This constitutes an improvement over \cite{Becher:2007cu}, where the factors $S(M_j^2)$ and $\widetilde Z(M_j^2)$ are absorbed into $S$ and $Z$. Doing so eliminates the separate RG evolution between thresholds and prevents the resummation of potentially large logarithms such as $\ln(m^2/M_j^2)$.

We now summarize the ingredients needed to obtain the two-loop results with multiple quarks. This includes:
\begin{itemize}
    \item in the hard function $C(Q^2)$, we now have to include $n_h + 1 + n_\ell$ massless quarks;
    \item the functions $S(M_j^2)$ and  $S(m_i^2)$ are simply given by \eqref{eq: soft result} with $m^2 \rightarrow M_j^2$ or $m^2 \rightarrow m_i^2$, since the effects from eq. \eqref{eq:coupling_replace} appear only at the three-loop level, and massless quark loops are scaleless;
    \item $\w Z(M_j^2)$ is calculated in Appendix \ref{app: wZ0} and it reads
    \begin{align} \notag
    \w Z(M_j^2) &= 1 + \left( \frac{\alpha_0}{4\pi} \right )^2  w^2(\nu) \left ( \frac{4\pi e^{-\gamma_E}}{M_j^2} \right )^{2\epsilon}\left(\frac{\nu^2}{Q^2}\right)^{\eta/2} C_F T_F \Bigg \{ \frac{8}{3\eta} (\Gamma(\epsilon) e^{\epsilon \gamma_E})^2 \frac{1+\epsilon}{1+ \frac{8}{3} \epsilon + \frac{4}{3} \epsilon^2}
\\ \label{eq: wZ0 twoloop}
&\quad + \frac{2}{\epsilon^2} + \frac{1}{\epsilon} \left ( - \frac{1}{3} - \frac{4\pi^2}{9} \right ) + \frac{73}{18} + \frac{29\pi^2}{27} - \frac{8}{3} \zeta_3 + \f O(\eta, \epsilon) \Bigg \} + \f O(\alpha_0^3);
    \end{align}
    \item $Z(m^2)$ is given by
\begin{align} \notag 
Z(m^2) &= 1 + \frac{\alpha_0}{4\pi} \left ( \frac{4\pi e^{-\gamma_E}}{m^2} \right )^{\epsilon} \Bigg \{ 1 - \frac{\alpha_0 T_F}{3\pi \epsilon} (4\pi)^{\epsilon} \Gamma(1+\epsilon) \sum_j (M_j^2)^{-\epsilon} \Bigg \} C_F Z^{(1)}
\\   \label{eq: Z0m2 twoloop} 
&\quad + \left ( \frac{\alpha_0}{4\pi} \right )^2 \left ( \frac{4\pi e^{-\gamma_E}}{m^2} \right )^{2\epsilon} w^2(\nu) \left ( \frac{\nu^2}{Q^2} \right )^{\eta/2} C_F \, \Bigg \{ C_F Z_F^{(2)} + C_A Z_A^{(2)}  
\\ \notag
&\qquad + T_F Z_T^{(2)} + T_Fn_\ell Z_l^{(2)} + \f O(\eta ; \epsilon) \Bigg \} + \f O(\alpha_0^3),
\end{align}
where $Z^{(1)}, Z_F^{(2)}, Z_A^{(2)}, Z_T^{(2)}$ are given in section \ref{sec: soft} and
\begin{align} \label{eq: Zl}
Z_l^{(2)} &= - \frac{2}{3 \epsilon^3} - \frac{16}{9\epsilon^2} + \left ( - \frac{149}{27} - \frac{5\pi^2}{9} \right ) \frac{1}{\epsilon}  - \frac{3269}{162} - \frac{40\pi^2}{27} - \frac{44}{9} \zeta_3;
\end{align}

\item $\w{\mathfrak Z}(m_i^2)$ is given by

\begin{align} \notag
  \w{\mathfrak Z}(m_i^2) &=   1+\left(\frac{\alpha_0}{4\pi}\right)^2 \left ( \frac{\nu^2}{m_i^2} \right )^{\eta/2} \left ( \frac{4\pi e^{-\gamma_E}}{m_i^2} \right )^{2\epsilon} C_F T_F \Bigg \{ \frac{8}{3\eta} (\Gamma(\epsilon) e^{\epsilon \gamma_E})^2 \frac{1+\epsilon}{1+ \frac{8}{3} \epsilon + \frac{4}{3} \epsilon^2} 
  \\
  &\quad - \frac{2}{\epsilon^3} + \frac{4}{9\epsilon^2} + \frac{1}{\epsilon} \left ( \frac{86}{27} - \frac{\pi^2}{3} \right ) - \frac{128}{9} + \frac{2\pi^2}{27} + 4 \zeta_3 \Bigg \} \,.
\end{align}

\end{itemize}

\section{Boson mass as IR regulator} \label{sec: gluon mass reg}

\subsection{Factorization}
\label{sec:gluonmassreg_fact}

To fully expose the IR region structure of the form factor, we must introduce an additional IR regulator, e.g.\ an off-shellness or a small gauge boson mass. Here we use a gluon mass regulator, implemented by the replacement $k^2 \to k^2-m_g^2$ in the denominator of gluon propagators. We assume $m_g \ll m$ and neglect $\f O(m_g^2/m^2)$ terms. Since a gauge-boson mass breaks gauge invariance in non-abelian gauge theories, this section should be understood in the abelian (QED) limit (we keep the QCD notation for continuity).

With the gluon mass regulator present, the discussion from section \ref{sec: region cascade general discussion} changes significantly in that the ultra-collinear contributions are no longer scaleless. Indeed, as will be seen in the following, the relevant ultra-collinear scaling reads
\begin{align} \label{eq: uc mg}
p_{uc} \sim \frac{m_g}{m} p_c \sim \left ( \frac{m_g Q}{m}, m_g , \frac{m m_g}{Q} \right ),
\end{align}
which agrees with the $uc$ scaling in section \ref{sec: region cascade general discussion} given the identification $m_g \sim \lambda^2 m = \frac{m^3}{Q^2}$, although this identification is not necessary and may be misleading. One should think of $m_g$ as setting the scale for the ultra-collinear degrees of freedom that was absent when regulating IR divergences with dimensional  regularization. The cascade of regions still gives scaleless contributions unless the virtuality of the modes is $m_g^2$. 
Otherwise, the ultra-collinear modes are still described by bHQET, and the discussion of section~\ref{sec: region cascade general discussion} applies. To summarize, we consider a sequence of EFTs
\begin{align} \label{eq: EFT seq}
{\rm QCD}(Q^2) \rightarrow {\rm SCET_I}(mQ) \rightarrow {\rm SCET_{II}}(m^2) \rightarrow {\rm bHQET}(m_g^2)\,,
\end{align}
leading to the factorization formula
\begin{align} \label{eq: new fact}
F_1(m^2, m_g^2, Q^2) = C(Q^2)\, S(m^2)\, Z(m^2)\, \mathfrak S(m_g^2)\, \mathfrak Z(m_g^2)
+ \f O\!\left(\frac{m^2}{Q^2},\frac{m_g^2}{m^2}\right).
\end{align}
Here $\mathfrak{Z}$ is the bHQET ultra-collinear function defined in eq.~\eqref{eq:ZUCdef}; 
it corresponds to the matching coefficient $\mathfrak{Z}_1$ of section~\ref{sec: region cascade general discussion}, 
now rendered non-trivial because $m_g$ supplies the physical scale that was absent in pure dimensional regularization.
The function $\mathfrak S$ describes the emergent \textit{ultra-soft$_1$} modes
\begin{align} \label{eq: us mg}
p_{us_1} \sim (m_g, m_g , m_g).
\end{align}
It is defined by
\begin{align} \label{eq: frak S def}
\mathfrak S(m_g^2) = \frac{1}{N_c} \text{tr} \bra 0 \mathfrak S_-^{\dagger}(0) \mathfrak S_+(0) \ket 0,
\end{align}
where
\begin{align}
\mathfrak S_{\pm}(x) &= \mathcal{P} \exp \left [ ig \int_{-\infty}^0 ds \, n_{\pm} \cdot A_{us_1}(x + s n_{\pm} ) \right ]
\end{align}
are Wilson lines made from ultra-soft$_1$ gluon fields $A_{us_1}$. Note that the ultra-soft$_1$ modes have been sparingly discussed in section \ref{sec: region cascade general discussion}. In the massless gluon case, they are scaleless and do not change the conclusion of section \ref{sec: region cascade general discussion}. With a non-zero gluon mass, they become non-zero. The relation between ultra-soft$_1$ and the ultra-(anti-)collinear modes, which live on the same invariant mass hyperbola, is of the ${\rm SCET_{II}}$-type. That is, $p_{uc} + p_{us_1}$ becomes off-shell and is to be integrated out. Since there are no ultra-soft$_1$ external particles, the ultra-soft$_1$ modes factorize into Wilson lines using standard arguments, resulting in the soft factor $\mathfrak S$. The corresponding decoupling transformation can be performed in the ${\rm SCET_{II}}(m^2)$ theory, before  matching onto ${\rm bHQET}(m_g^2)$. Indeed, the ultra-soft$_1$ modes behave in the same way as the ultra-soft modes in relation to the (anti-)collinear modes. 

Each function $S, Z, \mathfrak S, \mathfrak Z$ contains rapidity divergences that can be regulated by the $\eta$ regulator by modifying the definitions of the soft, collinear, ultra-soft$_1$, and ultra-collinear Wilson lines, respectively; see eq. \eqref{eq: eta regulator} and
\begin{align} \notag
\mathfrak S_{\pm}(x) &= \mathcal{ P} \exp\left ( i g_0 w(\nu) \nu^{\eta/2} \int_{-\infty}^0 ds\,  n_{\pm \mu} (|2 i \p^z|^{-\eta/2}A_{us}^{\mu})(x+sn_{\pm}) \right ),
\\ \label{eq: eta regulator 2}
W_{uc}(x) &= \mathcal{ P}  \exp\left ( i g_0 w^2(\nu) \nu^{\eta} \int_{-\infty}^0 ds\, n_{+\mu} (|in_+ \p|^{-\eta}A_{uc}^{\mu})(x+sn_+) \right ),
\\ \notag
W_{\widebar{uc}}(x) &= \mathcal{ P} \exp\left ( i g_0 w^2(\nu) \nu^{\eta} \int_{-\infty}^0 ds\, n_{-\mu} (|in_- \p|^{-\eta}A_{\widebar{uc}}^{\mu})(x+sn_-) \right ).
\end{align}
$S$ and $Z$ are the same functions as before, determined to two-loop accuracy in section \ref{sec: soft}. $\mathfrak S, \mathfrak Z$ are non-zero at one-loop and have rapidity divergences at this order. We will calculate those functions in section \ref{sec: mgamma oneloop}.

The quark on-shell wave-function renormalization constant $Z_2$ factorizes (see, for example,~\cite{Grozin:2010wa,Grozin:2020jvt})
\begin{align}
Z_2(m^2, m_g^2) = Z_2^{m_g = 0}(m^2) \, \mathfrak Z_2(m_g^2),
\end{align}
where
\begin{align}
\mathfrak Z_2 &= 1 + \frac{\alpha(\mu) C_F}{2\pi} \left ( \frac{1}{\epsilon} + \ln \frac{\mu^2}{m_g^2} + \f O(\epsilon) \right ) + \f O(\alpha^2)
\end{align}
is the bHQET (Wilson line) self energy.
Recall that $Z_2$ is part of $F_1$ through the LSZ prescription. Naturally, $Z_2^{m_g = 0}$ is to be absorbed into $Z$, and $\mathfrak Z_2$ into $\mathfrak Z$.

There are different ways in which the ultra-collinear and ultra-soft$_1$ modes become relevant. A different scale is obtained by introducing an off-shellness $p_{c,\bar c}^2 - m^2 \neq 0$. This was, in fact, discussed in \cite{Fleming:2007qr, Fleming:2007xt, Hoang:2015vua}, where the off-shellness is given by the difference of the jet invariant mass of a top-induced jet and the top mass. The IR regulator affects only the bHQET matrix elements, not those in SCET$_{\rm II}$. Hence, as we have already noted, the functions appearing in \cite{Hoang:2015vua} that describe modes at the top quark mass $m_t^2$ (which corresponds to our $m^2$) coincide with the functions $Z,S$ that we have calculated.

\subsection{One-loop calculation} \label{sec: mgamma oneloop}
The vertex correction to $F_1$ is given in terms of the integral
\begin{align}
I &= \frac{2\alpha_0}{(4\pi)^{1-\epsilon}} \int \frac{d^dk}{i\pi^{d/2}} \frac{ n_- (p_{\bar c} + k) \, n_+ (p_c + k)  }{[k^2 - m_g^2] [2p_{\bar c} k + k^2][2p_ck + k^2]}.
\end{align}
We have already determined that the hard region $k \sim (Q,Q,Q)$ gives
\begin{align}
I_h &= \frac{\alpha(\mu)}{2\pi}  \left \{  - \frac{1}{\epsilon^2} - \frac{1}{\epsilon} \left (\ln \frac{\mu^2}{Q^2} + \frac{3}{2}\right ) - \frac{1}{2} \ln^2 \frac{\mu^2}{Q^2} - \frac{3}{2}\ln \frac{\mu^2}{Q^2} - 4 + \frac{\pi^2}{12} \right \}.
\end{align}
and the collinear region $k \sim (Q, m^2/Q, m )$, giving
\begin{align}
I_c = \frac{\alpha(\mu)}{4\pi}  \Bigg \{ \frac{1}{\epsilon^2} + \frac{1}{\epsilon} \left ( \ln \frac{\mu^2}{m^2}  + 2 \right ) + \frac{1}{2 }\ln^2 \frac{\mu^2}{m^2}  + 2\ln \frac{\mu^2}{m^2} + 4 + \frac{\pi^2}{12} \Bigg \}.
\end{align}
The soft region $k \sim (m , m , m)$ gives a scaleless integral, i.e. the soft function $S$ is zero at one-loop, as before. However, the ultra-soft$_1$ contribution $k \sim (m_g, m_g, m_g)$ is non-scaleless
\begin{align} \notag
I_{us_1} &= \frac{2\alpha_0 w^2(\nu) \nu^{\eta} }{(4\pi)^{1-\epsilon}} \int \frac{d^dk}{i\pi^{d/2}} \frac{|2k^z|^{-\eta}}{[k^2 - m_g^2] [n_+ k ][n_- k ]}
\\
&= \frac{\alpha(\mu)}{2\pi}  \Bigg \{ - w^2(\nu) \frac{2}{\eta} \left ( \frac{\mu^2}{m_g^2} \right )^{\epsilon}  \Gamma(\epsilon) e^{\epsilon \gamma_E} + \frac{1}{\epsilon^2} + \frac{1}{\epsilon} \ln \frac{\mu^2}{\nu^2} 
\\ \notag
&\quad + \frac{1}{2} \ln^2 \frac{\mu^2}{m_g^2} - \ln \frac{\mu^2}{m_g^2} \ln \frac{\nu^2}{m_g^2} - \frac{\pi^2}{12}\Bigg \} + \f O(\eta, \epsilon),
\end{align}
where one may find the result for this simple integral in \cite{Chiu:2012ir}. The remaining contribution comes from the ultra-collinear region:
\begin{align} \notag
I_{uc} &= \frac{2\alpha_0 w^2(\nu) \nu^{\eta}}{(4\pi)^{1-\epsilon}} \int \frac{d^dk}{i\pi^{d/2}} \frac{|n_+ k|^{-\eta}}{[k^2 - m_g^2] [n_+ k ][v_-k]}
\\ \notag
&= - \frac{\alpha(\mu) w^2(\nu)}{4\pi} \, e^{\epsilon \gamma_E} \Gamma(\epsilon + \eta/2) \Gamma(-\eta/2) \left ( \frac{m\nu}{m_gQ} \right )^{\eta} \left( \frac{\mu^2}{m_g^2} \right )^{\epsilon}
\\
&= \frac{\alpha(\mu) }{4\pi} \, \Bigg \{ \frac{2}{\eta} w^2(\nu) e^{\epsilon \gamma_E} \Gamma(\epsilon) \left( \frac{\mu^2}{m_g^2} \right )^{\epsilon} - \frac{1}{\epsilon^2} - \frac{1}{\epsilon} \left ( \ln \frac{\mu^2}{m_g^2} - \ln \frac{m^2}{Q^2} - \ln \frac{\nu^2}{m_g^2} \right ) 
\\ \notag
&\quad - \frac{1}{2}  \ln^2 \frac{\mu^2}{m_g^2} + \ln \frac{\mu^2}{m_g^2} \left ( \ln \frac{m^2}{Q^2} + \ln \frac{\nu^2}{m_g^2} \right) + \frac{\pi^2}{12} + \f O(\eta, \epsilon) \Bigg \}.
\end{align}
Adding contributions from all the regions together gives
\begin{align} \notag
I &= I_h + 2 I_c + I_{us_1} + 2I_{uc}
\\
&= \frac{\alpha(\mu)}{4\pi} \Bigg \{ \frac{1}{\epsilon} - \ln^2 \frac{\mu^2}{Q^2} - 3\ln \frac{\mu^2}{Q^2} + \ln^2 \frac{\mu^2}{m^2}  + 4\ln \frac{\mu^2}{m^2}+ 2\ln \frac{\mu^2}{m_g^2} \ln \frac{m^2}{Q^2} + 4 + \frac{\pi^2}{3} + \f O(\eta, \epsilon) \Bigg \}
\end{align}
and adding the LSZ contributions gives
\begin{align}
F_1 &= 1 + \frac{\alpha_0 C_F}{2\pi} \Bigg \{- \ln \frac{m_g^2}{Q^2} \left ( \ln \frac{m^2}{Q^2} + 1 \right ) + \frac{1}{2} \ln^2 \frac{m^2}{Q^2} - \frac{1}{2} \ln \frac{m^2}{Q^2} + \frac{\pi^2}{6}  + \f O(\eta, \epsilon) \Bigg \} + \f O(\alpha_0^2).
\end{align}
We have also obtained
\begin{align} \notag
\mathfrak S \left (m_g^2, \mu, \frac{\nu}{m_g} \right ) &= 1 +  \frac{\alpha(\mu) C_F}{2\pi}  \Bigg \{ - w^2(\nu) \frac{2}{\eta} \left ( \frac{\mu^2}{m_g^2} \right )^{\epsilon}  \Gamma(\epsilon) e^{\epsilon \gamma_E} + \frac{1}{\epsilon^2} + \frac{1}{\epsilon} \left ( \ln \frac{\mu^2}{m_g^2} - \ln \frac{\nu^2}{m_g^2} \right ) 
\\  \label{eq: frak S oneloop result}
&\quad + \frac{1}{2} \ln^2 \frac{\mu^2}{m_g^2} - \ln \frac{\mu^2}{m_g^2} \ln \frac{\nu^2}{m_g^2} - \frac{\pi^2}{12} + \f O(\eta, \epsilon)\Bigg \}  + \f O(\alpha^2),
\\ \notag
\mathfrak Z\left (m_g^2, \mu, \frac{m\nu}{m_g Q} \right ) &= 1 + \frac{\alpha(\mu)C_F }{2\pi} \, \Bigg \{  w^2(\nu) \frac{2}{\eta} e^{\epsilon \gamma_E} \Gamma(\epsilon) \left( \frac{\mu^2}{m_g^2} \right )^{\epsilon} - \frac{1}{\epsilon^2} + \frac{1}{\epsilon} \left ( \ln \frac{m^2 \nu^2}{Q^2 m_g^2} - \ln \frac{\mu^2}{m_g^2} + 1 \right )
\\ 
&\quad - \frac{1}{2}  \ln^2 \frac{\mu^2}{m_g^2} + \ln \frac{\mu^2}{m_g^2} + \ln \frac{\mu^2}{m_g^2}  \ln \frac{m^2 \nu^2}{Q^2 m_g^2}  + \frac{\pi^2}{12} + \f O(\eta, \epsilon)  \Bigg \} + \f O(\alpha^2).
\end{align}

\subsection{Renormalization and resummation}
\label{sec:resum}
In this section, we restrict ourselves to one-loop accuracy, so we can ignore $S$ in eq.~\eqref{eq: new fact}; furthermore, $Z$ contains no rapidity divergences at this order. We define the renormalization constants ${\rm Z}_X$ and anomalous dimensions by subtracting all $\frac{1}{\epsilon}$ (UV and IR) and $\frac{1}{\eta}$ poles in $\ms$:
\begin{align}
X = {\rm Z}_X X_R, \qquad \gamma_X^{(\mu)} = {\rm Z}_X^{-1} \frac{d}{d\ln \mu} {\rm Z}_X, \qquad \gamma_X^{(\nu)} = {\rm Z}_X^{-1} \frac{d}{d\ln \nu} {\rm Z}_X.
\end{align}
For $X \in \{ C, Z, S \}$, the results were given in eq. \eqref{eq: ADs 1}. For $X \in \{ \mathfrak Z, \mathfrak S \}$, we get
\begin{align} 
\gamma_{\mathfrak S}^{(\mu)} &= -\frac{\alpha(\mu)}{\pi} C_F \, \ln \frac{\mu^2}{\nu^2} + \f O(\alpha^2),
\qquad
&&\gamma_{\mathfrak Z}^{(\mu)} = \frac{\alpha(\mu)}{\pi} C_F \left ( \ln \frac{\mu^2 Q^2}{m^2\nu^2 }    - 1 \right ) + \f O(\alpha^2)
\\ \notag
\gamma_{\mathfrak S}^{(\nu)} &= \frac{\alpha(\mu)}{\pi} C_F \ln \frac{\mu^2}{m_g^2} + \f O(\alpha^2), \qquad &&\gamma_{\mathfrak Z}^{(\nu)} = - \frac{\alpha(\mu)}{\pi} C_F \ln \frac{\mu^2}{m_g^2} + \f O(\alpha^2).
\end{align}
Following \cite{Chiu:2012ir}, we perform the resummation using the fixed-order form of $\gamma^{(\nu)}_X$ running in $\nu$ first and then in $\mu$. We have
\begin{align} \notag
C_R( Q^2 , \mu ) &= U_C(\mu, \mu_C) C_R(Q^2, \mu_C),
\\ \notag
Z_R( m^2, \mu) &= U_Z(\mu, \mu_Z) Z_R(m^2 , \mu_Z),
\\
\mathfrak S_R\left ( m_g^2, \mu , \frac{\nu}{m_g} \right ) &= V_{\mathfrak S}(\nu, \nu_{\mathfrak S} ; \mu) U_{\mathfrak S}(\mu , \mu_{\mathfrak S} ; \nu_{\mathfrak S}) \mathfrak S_R\left ( m_g^2, \mu_{\mathfrak S} , \frac{\nu_{\mathfrak S}}{m_g} \right ),
\\ \notag
\mathfrak Z_R\left ( m_g^2, \mu , \frac{m\nu}{m_g Q} \right ) &=  V_{\mathfrak Z}(\nu, \nu_{\mathfrak Z} ; \mu) U_{\mathfrak Z}(\mu , \mu_{\mathfrak Z} ; \nu_{\mathfrak Z})\mathfrak Z_R\left ( m_g^2, \mu_{\mathfrak Z} , \frac{m\nu_{\mathfrak Z}}{m_g Q} \right ),
\end{align}
where 
\begin{align}
\ln U_C(\mu, \mu_C) &= - \frac{8\pi C_F}{\beta_0^2} \left ( \frac{1}{\alpha(\mu)} - \frac{1}{\alpha(\mu_C)} + \frac{1}{\alpha(Q)} \ln \frac{\alpha(\mu)}{\alpha(\mu_C)} \right ) + \frac{3C_F}{\beta_0} \ln \frac{\alpha(\mu)}{\alpha(\mu_C)},
\\
\ln U_Z(\mu, \mu_Z) &= \frac{8\pi C_F}{\beta_0^2} \left ( \frac{1}{\alpha(\mu)} - \frac{1}{\alpha(\mu_Z)} + \frac{1}{\alpha(m)} \ln \frac{\alpha(\mu)}{\alpha(\mu_Z)} \right ) - \frac{C_F}{\beta_0} \ln \frac{\alpha(\mu)}{\alpha(\mu_Z)},
\\
\ln V_{\mathfrak S}(\nu, \nu_{\mathfrak S} ; \mu) &=\frac{8\pi C_F}{\beta_0^2} \left ( \frac{1}{\alpha(\nu)} - \frac{1}{\alpha(\nu_{\mathfrak S})} \right ) \ln \frac{\alpha(\mu)}{\alpha(m_g)},
\\
\ln V_{\mathfrak Z}(\nu, \nu_{\mathfrak Z} ; \mu) &=  - \frac{8\pi C_F}{\beta_0^2} \left ( \frac{1}{\alpha(\nu)} - \frac{1}{\alpha(\nu_{\mathfrak Z})} \right ) \ln \frac{\alpha(\mu)}{\alpha(m_g)} ,
\\
\ln U_{\mathfrak S}(\mu , \mu_{\mathfrak S} ; \nu_{\mathfrak S}) &= \frac{8\pi C_F}{\beta_0^2} \left ( \frac{1}{\alpha(\mu)} - \frac{1}{\alpha(\mu_{\mathfrak S})} + \frac{1}{\alpha(\nu_{\mathfrak S})} \ln \frac{\alpha(\mu)}{\alpha(\mu_{\mathfrak S})} \right ) ,
\\
\ln U_{\mathfrak Z}(\mu , \mu_{\mathfrak Z} ; \nu_{\mathfrak Z}) &= - \frac{8\pi C_F}{\beta_0^2} \left ( \frac{1}{\alpha(\mu)} - \frac{1}{\alpha(\mu_{\mathfrak Z})} + \frac{1}{\alpha( \frac{m \nu_{\mathfrak Z}}{Q})} \ln \frac{\alpha(\mu)}{\alpha(\mu_{\mathfrak Z})} \right ) - \frac{2C_F}{\beta_0} \ln \frac{\alpha(\mu)}{\alpha(\mu_{\mathfrak Z})}. 
\end{align}
Thus the full evolution factor reads
\begin{align} \notag
&U( \mu_C, \mu_Z, \mu_{\mathfrak Z}, \mu_{\mathfrak S} ; \nu_{\mathfrak Z}, \nu_{\mathfrak S}) = U_C(\mu, \mu_C) U_Z(\mu, \mu_Z) V_{\mathfrak S}(\nu, \nu_{\mathfrak S} ; \mu) U_{\mathfrak S}(\mu , \mu_{\mathfrak S} ; \nu_{\mathfrak S}) V_{\mathfrak Z}(\nu, \nu_{\mathfrak Z} ; \mu) U_{\mathfrak Z}(\mu , \mu_{\mathfrak Z} ; \nu_{\mathfrak Z})
\\ 
&\quad = \exp \Bigg \{ \frac{8\pi C_F}{\beta_0^2} \Bigg [ \frac{\ln \frac{\alpha(\mu_C)}{\alpha(\mu_{\mathfrak Z})}}{\alpha(Q)} - \frac{\ln \frac{ \alpha(\mu_Z) }{\alpha(\mu_{\mathfrak Z})}}{\alpha(m)} - \frac{\ln \frac{ \alpha(m_g) }{\alpha(\mu_{\mathfrak Z})}}{\alpha(\nu_{\mathfrak Z})}   + \frac{\ln \frac{\alpha(m_g)}{\alpha(\mu_{\mathfrak S})}}{\alpha(\nu_{\mathfrak S})}  
\\ \notag
&\qquad + \frac{1}{\alpha(\mu_C)} - \frac{1}{\alpha(\mu_Z)} + \frac{1}{\alpha(\mu_{\mathfrak Z})} - \frac{1}{\alpha(\mu_{\mathfrak S})} \Bigg ] + \frac{C_F}{\beta_0} \left [ - 3 \ln \alpha(\mu_C) + \ln \alpha(\mu_Z) + 2 \ln \alpha(\mu_{\mathfrak Z}) \right ] \Bigg \},
\end{align}
where we used the leading-logarithmic running, $\alpha^{-1}(\mu)\simeq\alpha^{-1}(Q)+\frac{\beta_0}{4\pi}\ln(\mu^2/Q^2)$, which gives
\begin{align}
\frac{1}{\alpha\!\left( \frac{m \nu_{\mathfrak Z}}{Q}\right)} = - \frac{1}{\alpha(Q)} + \frac{1}{\alpha(m)} + \frac{1}{\alpha(\nu_{\mathfrak Z})}
\end{align}
exactly within this approximation. For the scale choices such that no logarithmic corrections remain in the initial conditions
\begin{align} 
&\ln U\left ( Q, m, m_g, m_g ; \frac{m_g Q}{m}, m_g \right ) 
\\  \notag
&= \frac{8\pi C_F}{\beta_0^2} \Bigg [ \frac{1 + \ln \frac{\alpha(Q)}{\alpha(m_g)}}{\alpha(Q)} - \frac{1 + \ln \frac{ \alpha(m) }{\alpha(m_g)}}{\alpha(m)} \Bigg ] + \frac{C_F}{\beta_0} \left [ - 3 \ln \alpha(Q) + \ln \alpha(m) + 2 \ln \alpha(m_g) \right ]. 
\end{align}
This exponential contains all logarithms of $F_1$. The resummed IR divergent factor of $F_1$ at one loop is 
\begin{align}
F_1^{\rm IR, LO} =  \left ( \frac{ \alpha(\mu) }{\alpha(m_g)} \right )^{ -\frac{8\pi C_F}{\beta_0^2} \left ( \frac{1}{\alpha(m)} - \frac{1}{\alpha(Q)} \right ) - \frac{2C_F}{\beta_0} } \simeq \left ( \frac{ \alpha(\mu) }{\alpha(m_g)} \right )^{ - \frac{2C_F}{\beta_0} (1 + \ln \frac{m^2}{Q^2} ) },
\end{align}
where $\mu$ is some arbitrary scale.
The leading logarithmic solution
\begin{align}
\frac{\alpha(\mu)}{\alpha(m_g) } \simeq 1 + \frac{\alpha(\mu)\beta_0}{4\pi}  \ln \frac{m_g^2}{\mu^2}
\end{align}
gives
\begin{align}
F_1^{\rm IR, LO} \simeq \exp \left \{ - \frac{\alpha(\mu)C_F}{2\pi} \ln \frac{m_g^2 }{\mu^2} \left (1 + \ln \frac{m^2}{Q^2} \right )\right \}.
\end{align}
We remark on the well-known result that this becomes exact in the abelian case in the next section. In the non-abelian case, this fails already at two loops.

\subsection{Abelian exponentiation} \label{sec: abelian exp}
In this subsection, we specialize to the abelian theory. We therefore write the coupling and photon-mass regulator as $e$ and $m_{\gamma}$ (instead of $g$ and $m_g$) to make the restriction to QED explicit. The Sudakov form factor in QED and its resummation are classic results \cite{Sudakov:1954sw,Mueller:1979ih}; here we show that they follow directly from the factorized expressions obtained in section~\ref{sec:gluonmassreg_fact} once these results are given their consistent EFT interpretation.

The abelian bHQET, as well as the ultra-soft$_1$ effective theory, are described by the leading power Lagrangian:
\begin{align}
\f L_{uc} &= \bar h_{uc} i v_- D_{uc} \frac{\s n_+}{2} h_{uc} - \frac{1}{4} F_{uc \mu \nu} F_{uc}^{\mu \nu}\,;
\\
\f L_{us_1} &= - \frac{1}{4} F_{us_1 \mu \nu} F_{us_1}^{\mu \nu}\,,
\end{align}
defined at the scale $m_{\gamma} \ll m$. At leading power, the interactions in these sectors can be removed from the Lagrangian by standard field redefinitions. In particular, $\f L_{us_1}$ contains only the (regulated) photon field and therefore describes a non-interacting massive vector field with $m_\gamma$ implemented as in section~\ref{sec:gluonmassreg_fact}. In the ultra-collinear sector, the fermion couples eikonally through $v_- D_{uc}$, and we can eliminate this coupling by the field redefinition $h_{uc}(x) \rightarrow Y_{v_-}(x)\, h_{uc}(x)$, where
\begin{align}
Y_{v_-}(x) = \exp \left ( i \tilde e_0 \int_{-\infty}^0 ds\, v_- A_{uc}(x + sv_-) \right ).
\end{align}
After this, $h_{uc}$ is a sterile field with Lagrangian $\bar h_{uc} i v_- \partial \frac{\s n_+}{2} h_{uc}$.

We denote by $\tilde e_0$ the coupling constant appearing in the ultra-soft$_1$ and ultra-collinear Wilson lines. At leading power in the effective theory below the electron mass scale $m$, there are no dynamical charged degrees of freedom. As a result, the photon vacuum polarization vanishes in the low-energy theory, and $\tilde e_0$ is a finite, scale-independent parameter. 
In the EFT paradigm, $\tilde e_0$ is fixed by matching the low-energy theory to the UV theory, i.e., massive $\text{SCET}_{\text{II}}$, at the threshold scale $\mu \sim m$. This matching is directly analogous to decoupling a heavy flavor ($n_f=1 \to n_f=0$) in QCD. The matching condition requires
\begin{align} \label{eq: tilde alpha}
\frac{\tilde e_0^2}{4\pi} = \left ( \frac{\mu^2 e^{-\gamma_E}}{4\pi} \right )^{\epsilon} \alpha(\mu)  = Z_3 \alpha_0 \,,
\end{align}
where $\alpha_0$ is the \emph{bare} coupling constant of the massive theory and
\begin{align} \label{eq: Z3}
Z_3 = 1 - \frac{\alpha_0}{3\pi \epsilon} \Gamma(1+\epsilon) (4\pi)^{\epsilon} (m^2)^{-\epsilon} + \mathcal{O}(\alpha_0^2)
\end{align}
is the photon field renormalization constant in the on-shell scheme. By virtue of the QED Ward identity, $Z_3$ provides the complete renormalization of the coupling constant.

Since the ultra-soft$_1$ theory contains only the (regulated) photon field and the Wilson lines are abelian, the vacuum matrix element defining $\mathfrak S$ is governed by a Gaussian path integral. Equivalently, in perturbation theory only the connected two-point contraction contributes, and all higher-order terms are generated by exponentiating \cite{Yennie:1961ad} the one-loop exchange between the Wilson lines. Performing the Wick contractions in the ultra-soft$_1$ function to all orders, we thus find that $\mathfrak S$ is the exponential of its one-loop value, i.e.
\begin{align} \notag
\mathfrak S  &= \exp \Bigg \{ \frac{\alpha(\mu)}{2\pi}  \Bigg [ - w^2(\nu) \frac{2}{\eta} \left ( \frac{\mu^2}{m_{\gamma}^2} \right )^{\epsilon}  \Gamma(\epsilon) e^{\epsilon \gamma_E} + \frac{1}{\epsilon^2} + \frac{1}{\epsilon} \left ( \ln \frac{\mu^2}{m_{\gamma}^2} - \ln \frac{\nu^2}{m_{\gamma}^2} \right ) 
\\  \label{eq: S abelian allorder}
&\quad + \frac{1}{2} \ln^2 \frac{\mu^2}{m_{\gamma}^2} - \ln \frac{\mu^2}{m_{\gamma}^2} \ln \frac{\nu^2}{m_{\gamma}^2} - \frac{\pi^2}{12} + \f O(\eta, \epsilon) \Bigg ]\Bigg \}.
\end{align}

For the ultra-collinear function $\mathfrak Z$, the only extra bookkeeping concerns the self-energy of the $Y_{v_-}$ Wilson line. By definition, this contribution is absorbed into the wave-function factor $\mathfrak Z_2$, so that the matrix element entering $\mathfrak Z$ may be written as
\begin{align}
\sqrt{ \mathfrak Z } = \bra 0 W_{uc}^{\dagger}(0) Y_{v_-}(0) \ket 0 \, \times \sqrt{\mathfrak Z_2} = \bra 0 W_{uc}^{\dagger}(0) Y_{v_-}(0) \ket 0 |_{\text{no }Y_{v_-}\text{ self energy}}.
\end{align}
We note that this split is gauge dependent, and we restrict ourselves here to Feynman gauge. 
With the $Y_{v_-}$ self-energy removed in this way, the remaining ultra-collinear matrix element again reduces to Wick contractions of a free abelian gauge field between Wilson lines, and therefore exponentiates. Performing the Wick contractions for $\mathfrak Z$ to all orders, we obtain
\begin{align} \notag
\mathfrak Z&= \exp \Bigg \{ \frac{\alpha(\mu)}{2\pi} \, \Bigg [  w^2(\nu) \frac{2}{\eta} e^{\epsilon \gamma_E} \Gamma(\epsilon) \left( \frac{\mu^2}{m_{\gamma}^2} \right )^{\epsilon} - \frac{1}{\epsilon^2} + \frac{1}{\epsilon} \left ( \ln \frac{m^2 \nu^2}{Q^2 m_{\gamma}^2} - \ln \frac{\mu^2}{m_{\gamma}^2} + 1 \right )
\\ 
&\quad - \frac{1}{2}  \ln^2 \frac{\mu^2}{m_{\gamma}^2} + \ln \frac{\mu^2}{m_{\gamma}^2} + \ln \frac{\mu^2}{m_{\gamma}^2}  \ln \frac{m^2 \nu^2}{Q^2 m_{\gamma}^2}  + \frac{\pi^2}{12} + \f O(\eta, \epsilon)  \Bigg ] \Bigg \}.
\end{align}

Multiplying the all-order expressions for $\mathfrak Z$ and $\mathfrak S$ in the abelian theory, the rapidity-singular terms proportional to $w^2(\nu)/\eta$ cancel, and the dependence on the auxiliary rapidity scale $\nu$ drops out, as it must for the physical form factor. The infrared-sensitive factor, therefore, exponentiates to the familiar QED result,
\begin{align} \label{eq: F1IR 1}
F_1^{\rm IR} = \mathfrak Z\left (m_{\gamma}^2, \mu, \frac{m\nu}{Qm_{\gamma}} \right ) \, \mathfrak S \left (m_{\gamma}^2, \mu, \frac{\nu}{m_{\gamma}} \right ) =\exp \left \{ - \frac{\alpha(\mu)}{2\pi} \ln \frac{m_{\gamma}^2 }{\mu^2} \left (1 + \ln \frac{m^2}{Q^2} \right )\right \}
\end{align}
with a photon mass regulator.

With $m_{\gamma}=0$ in pure dimensional regularization, the same abelian exponentiation is captured entirely by the multiplicative counterterms: the infrared singular factor can be written as the inverse product of the renormalization factors introduced above, or equivalently as the product of the corresponding $Z$-factors for the ultra-soft$_1$ and ultra-collinear functions,
\begin{align} \label{eq: F1IR 2}
F_1^{{\rm IR}, m_{\gamma} = 0} = {\rm Z}_C {\rm Z}_S {\rm Z}_{Z} = \frac{1}{{\rm Z}_{\mathfrak S} {\rm Z}_{\mathfrak Z}} = \exp \left \{ \frac{\alpha(\mu)}{2\pi} \frac{1}{\epsilon} \left ( \ln \frac{m^2}{Q^2} + 1 \right ) \right \}.
\end{align}

In both eqs.~\eqref{eq: F1IR 1} and \eqref{eq: F1IR 2}, the scale $\mu$ plays the role of an arbitrary subtraction (or factorization) scale at which the infrared divergences are removed from the finite remainder. Accordingly, we may write the form factor as
\begin{align}
F_1 = F_1^{\rm fin}(\mu) F_1^{\rm IR}(\mu),
\end{align}
where $F_1^{\rm fin}(\mu)$ is finite and the $\mu$ dependence cancels between the two factors in the product.

Equations~\eqref{eq: F1IR 1} and \eqref{eq: F1IR 2} are the amplitude-level statements of abelian exponentiation for the Sudakov form factor. In particular, the large kinematic logarithm $\ln(m^2/Q^2)$ appears already in the exponent and is therefore resummed in the strict sense, such that expanding \eqref{eq: F1IR 1} generates the entire tower of terms
$\alpha^n \ln^n(m_{\gamma}^2/\mu^2)\,\bigl(1+\ln(m^2/Q^2)\bigr)^n/n!$.
From the EFT point of view, this logarithm is tied to the rapidity separation between the ultra-(anti-)collinear and ultra-soft$_1$ sectors: one may equivalently organize its resummation through rapidity evolution in the auxiliary scale $\nu$ (rapidity renormalization group), however in QED, the all-order result collapses to the one-loop exponent displayed above.

This simplification is specific to the abelian theory. In a non-abelian gauge theory, the corresponding correlators of Wilson lines are not Gaussian, and the logarithm of the soft factor receives contributions from connected ``webs'' with non-trivial color factors. The resulting resummation is then governed by the cusp anomalous dimension rather than by a simple exponentiation of the one-loop diagram; see e.g.\ \cite{Gatheral:1983cz,Frenkel:1984pz,Mitov:2010rp}.

Finally, the virtual factor $F_1^{\rm IR}$ is an infrared divergent quantity. For an inclusive quantity that sums over unobserved soft-photon emissions below a resolution scale, the real-emission contributions involve the same eikonal Wilson lines and also exponentiate. The soft infrared singularities (the $1/\epsilon$ poles in pure dimensional regularization, or the $\ln m_{\gamma}$ terms with a photon-mass regulator) cancel between virtual corrections and the phase-space integrals over real radiation, in accordance with the Bloch--Nordsieck/KLN mechanism and in direct correspondence with the standard Yennie-Frautschi-Suura organization of soft-photon effects \cite{Bloch:1937pw,Yennie:1961ad,Kinoshita:1962ur,Lee:1964is}.

To exhibit the cancellation of the infrared singularities in $F_1$ in a manifestly EFT way, we need to embed the form factor into a sufficiently inclusive observable in which one sums over unobserved soft and ultra-collinear radiation. In SCET/bHQET, real soft-photon emission is encoded in a measurement-dependent soft function defined as the vacuum matrix element of an operator built from soft and soft-collinear Wilson lines that appear in the factorized amplitude. The precise structure depends on the parametric scaling of the resolution variable $\Delta E$, for the unobserved radiation, which is defined after squaring the amplitude; see, for example, \cite{Fontes:2024yvw, Fontes:2025mps,Fontes:2025xbt}, where the soft and ultra-collinear functions contribute explicitly.

\section{Scalar gluon form factor} \label{sec: gluon FF}

So far, we have discussed the fermion form factor. In this section, we consider the scalar gluon form factor in the presence of a small gluon-mass regulator $m_g$ and its leading-power factorization, assuming hierarchy
$Q^2 \gg m^2 \gg m_g^2$.
We define the corresponding form factor by
\begin{equation} \label{eq: F1g def}
F_1^g= \frac{1}{2(1-\epsilon) p_c p_{\bar c}} \left ( - g_{\mu \nu} + \frac{p_{c\mu} p_{\bar c\nu} + p_{c\nu} p_{\bar c\mu}}{p_cp_{\bar c}} \right ) \frac{\delta^{ab}}{N_c} \Gamma_g^{\mu\nu ab}\,,
\end{equation}
where $\Gamma^{\mu\nu ab}_g$ is the effective vertex for a color-singlet scalar current coupling to two external gluons with momenta $p_c$ and $p_{\bar c}$, with $2p_c p_{\bar c}=Q^2$. Explicitly,
\begin{align}
\varepsilon_{\mu} \varepsilon_{\nu}' \Gamma_g^{\mu \nu ab} = - \frac{1}{4} \bra{g^b(p_{\bar c}, \varepsilon')}  F^{\rho \sigma c} F_{\rho \sigma}^c \ket{g^a(p_c, \varepsilon)}.
\end{align}
The external gluon states are defined by LSZ reduction; accordingly, each external leg contributes a factor $\sqrt{{\rm Z}_3}$, where
\begin{align} \notag
{\rm Z}_3 &= 1 + \frac{\alpha_0 }{4\pi } \, e^{\epsilon \gamma_E}\Gamma(1+\epsilon) \left ( \frac{4\pi e^{-\gamma_E}}{m^2} \right )^{\epsilon} T_F \left ( - \frac{4}{3\epsilon} \right ) + \left \{ \frac{\alpha_0}{4\pi}  e^{\epsilon \gamma_E}\Gamma(1+\epsilon) \left ( \frac{4\pi e^{-\gamma_E}}{m^2} \right )^{\epsilon} \right \}^2  T_F 
\\
&\quad \times \left \{ C_A \left ( - \frac{1}{\epsilon^2} - \frac{5}{2 \epsilon} + \frac{13}{12} \right ) + C_F \left ( - \frac{2}{\epsilon} - 15 \right ) + T_F \left ( - \frac{4}{3\epsilon} \right )^2 + \f O(\epsilon) \right \}
\end{align}
is the gluon wave-function renormalization constant in the on-shell scheme, i.e.\ ${\rm Z}_3 = \frac{1}{1-\pi(0)}$, with
\begin{align}
\pi^{\mu \nu}(k) = \left ( g^{\mu \nu} - \frac{k^{\mu}k^{\nu}}{k^2} \right ) \pi(k^2)
\end{align}
being the 1PI gluon self-energy.

Unlike \cite{Wang:2023qbf}, we do not renormalize the scalar gluon operator. This choice is purely a matter of bookkeeping: our goal is to determine the (massive-loop) gluon jet function, which is an infrared object and must not depend on how the hard current is renormalized. The two-loop expression for $F_1^g$ can be obtained from \cite{Wang:2023qbf} and \cite{Lee:2022nhh}. The result is given in appendix~\ref{app:F1NNLO}.

We now turn to the factorization of $F_1^g$. At leading power in $m^2/Q^2$, all dependence on the external partonic channel is confined to the hard matching of the QCD operator onto SCET$_{\rm II}$ operators; the subsequent mode separation and overlap subtractions are universal.\footnote{We only use a gluon mass $m_g$ as an infrared regulator. Since this obscures gauge invariance, we interpret $m_g$-dependent intermediate expressions accordingly and ultimately take the limit $m_g\to 0$.}
Proceeding as in the quark case, and taking $m\gg m_g$, we obtain the leading-power factorization
\begin{equation}\label{eq:F1gfactorization}
F_1^g
= C_g(Q^2)\, Z_g(m^2)\, S_g(m^2)\,
Z_g^{(m_g)}(m_g^2)\, S_g^{(m_g)}(m_g^2)
+ \mathcal{O}\!\left(\frac{m^2}{Q^2},\frac{m_g^2}{m^2}\right).
\end{equation}
Here $C_g(Q^2)$ is the hard matching coefficient (known up to four-loop order \cite{Lee:2022nhh}), and $Z_g(m^2)$ and $S_g(m^2)$ encode the heavy-fermion loop effects at the scale $m$ in the collinear and soft sectors, respectively. The additional factors $Z_g^{(m_g)}(m_g^2)$ and $S_g^{(m_g)}(m_g^2)$ are purely infrared objects in the low-energy (fermionless) theory and contain the dependence on the gluon-mass regulator. In pure dimensional regularization ($m_g=0$), these infrared factors are absent at the level of bare matrix elements (they are scaleless), and their effect is encoded in the usual $1/\epsilon$ infrared singularities; for the applications to massification discussed below, this is the relevant case.

The soft functions $S_g$ and $S_g^{(m_g)}$ are given by the same Wilson line correlators as in the quark form factor, see eqs.~\eqref{eq:Sdef} and \eqref{eq: frak S def}, but with Wilson lines in the adjoint representation. The corresponding subleading-order expressions are obtained from eqs.~\eqref{eq: soft result} and \eqref{eq: frak S oneloop result} by the replacement $C_F\to C_A$, respectively. The collinear and low-scale collinear functions are defined by
\begin{align} \label{eq: Zg def}
        \sqrt{Z_g(m^2)}\epsilon^\mu &= \langle0| W_c^\dagger(iD_{c\perp}^\mu W_c) |g_c\rangle,
        \\
        \sqrt{Z_g^{(m_g)}(m_g^2)} \epsilon^{\mu} &= \langle0|W_{uc}^\dagger(iD_{uc \perp}^\mu W_{uc})|g_c\rangle.
\end{align}
The first matrix element defines the gluon jet function with massless gluons and massive fermion loops. The second is the corresponding jet function in the low-energy theory below $\mu\sim m$, regulated by $m_g$ and with no dynamical fermions.\footnote{The labels ``$uc$'' and the use of $W_{uc}$ and $D_{uc}$ follow the conventions of Sec.~\ref{sec: gluon mass reg} and simply indicate collinear scaling at virtuality $m_g^2$.}

In the abelian case with no charged light fields, the low-energy theory becomes non-interacting, and all soft/collinear factors trivialize, $Z_g^{(m_g)}=S_g^{(m_g)}=1$. The factorization formula for $m_g = 0$, i.e.\ with the IR divergences regularized dimensionally, is obtained from eq.~\eqref{eq:F1gfactorization} by setting the bare functions $Z_g^{(m_g)}=S_g^{(m_g)}=1$.

The massive gluon jet function $Z_g$ has been computed in \cite{Wang:2023qbf} up to two loops. However, there the rapidity logarithms were attributed solely to the soft function by means of effectively performing a soft subtraction, giving rise to logarithms $\ln\frac{Q^2}{m^2}$ in the soft function. Since this is disadvantageous for the resummation of large logarithms, we employ a different approach here and compute the collinear function with a rapidity regulator, without any subtractions. As for the heavy quark case, we can then employ rapidity RGEs to correctly resum the rapidity logarithms in the form factor. Using the same strategy as for $Z_q$, i.e. obtaining $Z_g$ from quantities computed in \cite{Wang:2023qbf}:
\begin{align}
Z_g = \frac{F_1^g}{C_g \, S_g} = \frac{\f Z_g^{} \f S^{}}{S_g},
\end{align}
where $\f Z_g$ has been computed in eqs. (2.6)-(2.8) of \cite{Wang:2023qbf} (taking $n_\ell = 0$ and $n_h = 1$) and $\f S$ is given in eqs. (2.3), (2.4) of \cite{Wang:2023qbf}.

Our result for $Z_g$, extracted from the NNLO expression of \cite{Wang:2023qbf}
using the strategy of section~\ref{sec: soft}, reads:
\begin{equation}
\begin{aligned} \label{eq: Zg result}
    Z_g(m^2)=&1+\frac{\alpha_0}{4\pi}\left ( \frac{4\pi e^{-\gamma_E}}{m^2} \right )^{\epsilon}     Z_g^{(1)}+\left(\frac{\alpha_0}{4\pi}\right)^2\left ( \frac{4\pi e^{-\gamma_E}}{m^2} \right )^{2\epsilon}w^2(\nu)\left(\frac{\nu^2}{Q^2}\right)^{\frac\eta2}Z_g^{(2)}+\mathcal{O}(\alpha^3),
    \end{aligned}
    \end{equation}
    with
    \begin{equation}
    \begin{aligned}    
    Z_g^{(1)}=&T_F\left ( - \frac{4}3 \right )\Gamma(\epsilon)e^{\epsilon \gamma_E},\\
    Z_g^{(2)} =& \, T_F C_A \frac{8}{3\eta}(\Gamma(\epsilon)e^{\epsilon\gamma_E})^2
   \frac{1+\epsilon}{1+\frac{8}{3}\epsilon+\frac{4}{3}\epsilon^2}
   + T_F C_A\left(-\frac{8}{3\epsilon} + \frac{10}{9} - 8\zeta_3\right)\\
&\quad + T_F C_F \left ( - \frac{2}{\epsilon} - 15 \right )  + T_F^2 \left ( \frac{16}{9\epsilon^2} + \frac{8\pi^2}{27} \right ) + \f O(\eta,\epsilon).
    \end{aligned}
\end{equation}
The anomalous dimension can be directly extracted from the single poles in the above expressions.

\section{Summary and outlook}
\label{sec:conclusion}

The Sudakov form factor with massive external fermions is the canonical testing ground for small-mass factorization. Beyond the standard hard, (anti-)collinear, and soft regions, individual multi-loop diagrams exhibit additional ultra-(anti-)collinear momentum regions with parametrically smaller virtualities. A long-standing question is whether these regions require extra factorization ingredients or whether they are an artifact of diagram-by-diagram expansions.

Our first result is that the entire ultra-collinear ``cascade'' cancels in the on-shell limit to all orders in perturbation theory because of gauge invariance. Furthermore, this cancellation holds even before considering inclusive observables.
In EFT language, this cancellation means that matching SCET$_{\rm II}$ onto the corresponding tower of boosted HQET descriptions yields bare matching functions that are scaleless on shell and therefore equal to unity. The standard leading-power factorization formula is thus not modified by ultra-(anti-)collinear messenger modes, even though those modes can appear in a regions analysis of scalar integrals and individual graphs in covariant gauge.

Our second result concerns massification (IR matching) \cite{Penin:2005eh,Mitov:2006xs,Becher:2007cu,Engel:2018fsb} in the boosted regime $Q^2\gg m^2$. At leading power, the full massive form factor can be written as a hard coefficient times \emph{universal} mass-dependent factors that carry the entire dependence on $\ln(m^2/Q^2)$. Using the $\eta$ rapidity regulator, we compute the relevant soft and (anti-)collinear functions through two loops and discuss their renormalization and combined $\mu$- and $\nu$-evolution. In this scheme, rapidity logarithms are resummed systematically by rapidity RG evolution, and the same universal building blocks can be extended directly to multi-leg amplitudes, including the case of additional heavy flavors and hierarchies of quark masses.

Dimensional regularization hides the physical content of scaleless regions. To make the infrared structure explicit, we also introduce an auxiliary gauge-boson mass as an IR regulator. This supplies a physical scale for the ultra-collinear and ultra-soft modes and makes their factorization manifest at the level of component functions. In this language, the well-known exponentiation of QED infrared divergences is an immediate corollary of EFT factorization.

Although the factors in \eqref{eq:fact1} are often referred to as ``single-scale'' SCET$_{\rm II}$ objects with a virtuality of order $m^2$, this is slightly misleading: as on-shell matrix elements, they still retain the infrared singularities of the form factor (and, in SCET$_{\rm II}$, rapidity/overlap singularities), so they are not purely short-distance quantities. 
In other words, the heart of the factorization problem is the separation of infrared poles into contributions that, once a physical measurement or real-emission contribution is specified, become logarithms of physical scales rather than remaining bare $1/\epsilon$ IR singularities. This necessitates a further low-energy matching below $m^2$ onto an EFT with messenger (soft-collinear/ultra-collinear) modes, i.e.\ bHQET, which disentangles the infrared content systematically and thereby forces the massive jet functions $Z_{c,\bar c}$ (and the accompanying soft factor) to be interpreted as \emph{matching coefficients} rather than bona fide matrix elements, even though the ultra-(anti-)collinear contributions cancel in the strictly on-shell limit.

Several directions for extension are natural. First, the present analysis is restricted to leading power in $m^2/Q^2$. Extending the same EFT logic to subleading power should make it unambiguous which additional operator structures and mode couplings are truly required to reproduce the mass-suppressed logarithms, and which apparently ``exotic'' regions disappear once gauge invariance and consistent overlap subtractions are enforced. While some steps have been made in this direction \cite{Penin:2014msa,vanBijleveld:2025ekz}, the results have not yet been phrased in a fully systematic EFT language.

A particularly timely application of such a subleading-power program is to gluon-induced, loop-mediated amplitudes, where next-to-leading-power (mass-suppressed) logarithms are both phenomenologically important and conceptually subtle due to endpoint/rapidity singularities and soft-fermion exchange contributions \cite{Wang:2019mym,Liu:2019oav,Liu:2021chn,Liu:2017vkm,Liu:2022ajh,Jaskiewicz:2024xkd}. In parallel, the current framework is well suited for systematizing ``massification'' beyond two loops and for processes with multiple scales and multiple fermion masses, where rapidity logarithms and possible factorization anomalies become central.

A related open problem concerns the case of multiple collinear directions in QCD, that is, amplitudes represented in SCET by an $N$-jet operator. In this case, the soft functions acquire non-trivial color structures. A detailed investigation of the factorization properties of the massive $N$-jet operator \cite{Becher:2009kw,Ferroglia:2009ii,Ferroglia:2009ep} in theories with multiple mass scales is left for future work.

\section*{Acknowledgements}
We thank Martin Beneke and Robin van Bijleveld for their valuable comments and discussions. This work was supported by the U.S. Department of Energy under Contract No. DE-SC0012704. J.S. was also supported by Laboratory Directed Research and Development (LDRD) funds from Brookhaven Science Associates. R.S. acknowledges the hospitality of the Kavli Institute for Theoretical Physics (KITP) during the completion of this work, supported by the National Science Foundation under Grant No. PHY-2309135.

\appendix

\section{$F_1$ and $F_1^g$ at NNLO} 
\label{app:F1NNLO}
We present the NNLO expression for $F_1$, extracted from \cite{Bernreuther:2006vp} (the $\sim \epsilon^2$ of the one-loop contribution is taken from \cite{Becher:2007cu}), in terms of the pole mass $m$ and the bare coupling constant $\alpha_0 = \frac{g_0^2}{4\pi}$. We have added the terms corresponding to heavier quarks according to section \ref{sec: more quarks}.
\begin{align}  \notag
F_1 &= 1 + \frac{\alpha_0}{4\pi} \left ( \frac{4\pi e^{-\gamma_E}}{m^2} \right )^{\epsilon}  C_F \f F_F^{(1)} + \left ( \frac{\alpha_0}{4\pi} \right )^2 \left ( \frac{4\pi e^{-\gamma_E}}{m^2} \right )^{2\epsilon} C_F \,
\\
&\quad \times \Bigg [ C_F \f F_F^{(2)} + C_A  \f F_A^{(2)}  + T_F \Bigg ( \f F_m^{(2)} + n_\ell \f F_l^{(2)} + \sum_j \f F_{M_j}^{(2)} \Bigg ) \Bigg ] + \f O(\alpha_0^3),
\end{align}
where 
\begin{align} \notag
\f F_F^{(1)} &= - \frac{2}{\epsilon} \left ( 1 + L \right ) - L^2 - 3 L - 4 + \frac{\pi^2}{3} + \epsilon \left [ - \frac{1}{3} L^3 - \frac{3}{2} L^2 + \left ( - 8 + \frac{\pi^2}{6} \right ) L - 8 + \frac{\pi^2}{3} + 4 \zeta_3 \right ]
\\ \notag
&\quad \quad + \epsilon^2 \Bigg [ - \frac{1}{12} L^4 - \frac{1}{2} L^3 + \left ( - 4 + \frac{\pi^2}{12} \right ) L^2 + \left ( - 16 + \frac{\pi^2}{4} + \frac{14}{3} \zeta_3 \right ) L 
\\
&\quad \quad  - 16 + \pi^2 + \frac{20}{3} \zeta_3 + \frac{7\pi^4}{90} \Bigg ] + \f O(\epsilon^3),
\\\notag
\f F_F^{(2)} &= \frac{2}{\epsilon^2} (L+1)^2 + \frac{1}{\epsilon} \left [ 2 L^3 + 8 L^2 + \left ( 14 - \frac{2\pi^2}{3} \right ) L + 8 - \frac{2\pi^2}{3} \right ]
\\ \notag
&\quad \quad + \frac{7}{6} L^4 + \frac{20}{3} L^3 + \left ( \frac{55}{2} - \frac{2\pi^2}{3} \right ) L^2  + \left ( \frac{85}{2} - 32 \zeta_3 \right ) L
\\
&\quad \quad  + 46 + \frac{13\pi^2}{2} - 44 \zeta_3 - \frac{59\pi^4}{90} - 8 \pi^2 \ln 2 + \f O(\epsilon),
\\ \notag
\f F_A^{(2)} &= -\frac{11}{3\epsilon^2} (L + 1) + \frac{1}{\epsilon} \left [ - \frac{11}{3} L^2 + \left ( - \frac{166}{9} + \frac{\pi^2}{3} \right ) L - \frac{181}{9} + \frac{14\pi^2}{9} - 2 \zeta_3 \right ] 
\\ \notag
&\quad \quad - \frac{22}{9} L^3 + \left ( - \frac{166}{9} + \frac{\pi^2}{3} \right ) L^2  + \left ( - \frac{4129}{54} - \frac{11 \pi^2}{18} + 26 \zeta_3 \right ) L 
\\
&\quad \quad - \frac{2387}{27} + \frac{59\pi^2}{54} + \frac{178}{3} \zeta_3 - \frac{\pi^4}{60} + 4\pi^2 \ln 2 + \f O(\epsilon), 
\\ \notag
\f F_m^{(2)} &= \frac{8}{3\epsilon^2} (L+1) + \frac{1}{\epsilon} \left ( \frac{4}{3} L^2 + 4 L + \frac{16}{3} - \frac{4\pi^2}{9} \right ) + \frac{8}{9} L^3  
\\
&\quad \quad + \frac{56}{9} L^2 + \left ( \frac{818}{27} + \frac{4\pi^2}{9} \right ) L + \frac{1820}{27} - \frac{8\pi^2}{9} - \frac{16}{3} \zeta_3 + \f O(\epsilon),
\\ \notag
\f F_l^{(2)} &= \frac{4}{3\epsilon^2} (L+1) + \frac{1}{\epsilon} \left ( \frac{4}{3} L^2 + \frac{56}{9} L + \frac{68}{9} - \frac{4\pi^2}{9} \right ) + \frac{8}{9} L^3 + \frac{56}{9} L^2
\\
&\quad \quad + \left ( \frac{706}{27} + \frac{2\pi^2}{9}  \right ) L + \frac{712}{27} - \frac{26\pi^2}{27} - \frac{32}{3} \zeta_3 + \f O(\epsilon),
\\ \notag
\f F_M^{(2)} &= \frac{8}{3\epsilon^2} (L+1) + \frac{1}{\epsilon} \left [ \frac{4}{3} L^2 + \left ( 4 + \frac{8}{3} L_M \right ) L + \frac{16}{3} - \frac{4\pi^2}{9} \right ] + \frac{8}{9} L^3 + \frac{56}{9}  L^2 
\\ \notag
&\quad \quad + \left ( \frac{8}{3} L_M^2 - \frac{40}{9} L_M + \frac{818}{27} + \frac{4\pi^2}{9} \right ) L - \frac{4}{9} L_M^3 + \frac{50}{9} L_M^2 + \left ( - \frac{386}{27} - \frac{8\pi^2}{9} \right ) L_M
\\ \label{eq: FM}
&\quad \quad + \frac{4483}{81} + \frac{32\pi^2}{27} - \frac{8}{3} \zeta_3 + \f O(\epsilon),
\end{align}
where $L = \ln \frac{m^2}{Q^2}, L_M = \ln \frac{m^2}{M^2}$.

The gluon form factor, including massive quarks with masses $m_1, m_2, ...$ and $n_\ell$ massless quarks, extracted from \cite{Wang:2023qbf} and \cite{Lee:2022nhh}, reads
\begin{align} \notag
F_1^g &= 1 + \frac{\alpha_0}{4\pi} \sum_f \left ( \frac{4\pi e^{-\gamma_E}}{m_f^2} \right )^{\epsilon} ( C_A \f F_A^{g(1)} + T_F \f F_T^{g(1)} )
\\ \notag
&\quad + \left ( \frac{\alpha_0}{4\pi} \right )^2 \sum_f \left ( \frac{4\pi e^{-\gamma_E}}{m_f^2} \right )^{2\epsilon}  \Big [ C_A^2 \, \f F_A^{g(2)} + T_F C_F \, \f F_F^{g(2)} + T_F C_A \f F_{TA}^{g(2)} + T_F^2 \, \f F_T^{g(2)} \Big ]
\\
&\quad + \left ( \frac{\alpha_0}{4\pi} \right )^2\left ( \frac{4\pi e^{-\gamma_E}}{Q^2} \right )^{2\epsilon} n_\ell T_F \left [ C_A \f F_{lA}^{g(2)} + C_F \f F_{lF}^{g(2)} \right ].
\end{align}
where $\sum_f$ goes over all massive quarks. Writing $L = \ln \frac{m_f^2}{Q^2}$, we find:
\begin{align} \notag
\f F_A^{g(1)} &= - \frac{2}{\epsilon^2} - \frac{2}{\epsilon} L - L^2 + \frac{\pi^2}{6} + \epsilon \left ( - \frac{1}{3} L^3 + \frac{\pi^2}{6} L - 2 + \frac{14}{3} \zeta_3 \right )
\\
&\quad + \epsilon^2 \left [ - \frac{1}{12} L^4 + \frac{\pi^2}{12} L^2 + \left ( - 2 + \frac{14}{3} \zeta_3 \right ) L - 6 + \frac{47\pi^4}{720} \right ] + \f O(\epsilon^3),
\\
\f F_T^{g(1)} &= - \frac{4}{3\epsilon} - \epsilon \frac{\pi^2}{9} + \epsilon^2 \frac{4}{9} \zeta_3 + \f O(\epsilon^3),
\\ \notag
\f F_A^{g(2)} &= \frac{2}{\epsilon^4} + \frac{1}{\epsilon^3} \left (4 L - \frac{11}{6} \right ) + \frac{1}{\epsilon^2} \left ( 4L^2 - \frac{11}{3} L - \frac{67}{18} - \frac{\pi^2}{6} \right ) 
\\ \notag
&\quad + \frac{1}{\epsilon} \left [ \frac{8}{3} L^3 - \frac{11}{3} L^2 + L \left (  - \frac{67}{9} - \frac{\pi^2}{3} \right ) + \frac{68}{27} + \frac{11\pi^2}{12} - \frac{25}{3} \zeta_3 \right ]
\\ \notag
&\quad + \frac{4}{3} L^4 - \frac{22}{9} L^3 + L^2 \left ( - \frac{67}{9} - \frac{\pi^2}{3} \right ) + L \left ( \frac{136}{27} + \frac{11\pi^2}{6} - \frac{50}{3} \zeta_3 \right )
\\
&\quad + \frac{5861}{162} + \frac{67\pi^2}{36} + \frac{11}{9} \zeta_3 - \frac{7\pi^4}{60} + \f O(\epsilon),
\\
\f F_F^{g(2)} &= - \frac{4}{\epsilon} - 4 L - \frac{112}{3} + 16 \zeta_3 + \f O(\epsilon),
\\ \notag
\f F_{TA}^{g(2)} &= \frac{16}{3\epsilon^3} + \frac{16}{3\epsilon^2} L + \frac{1}{\epsilon} \left ( \frac{8}{3} L^2 - \frac{20}{3} \right ) 
\\
&\quad + \frac{4}{3} L^3 + \frac{20}{3} L^2 + L \left ( \frac{8}{27} - \frac{4\pi^2}{9} \right ) - \frac{326}{81} - \frac{20\pi^2}{27} - \frac{248}{9} \zeta_3 + \f O(\epsilon),
\\
\f F_T^{g(2)} &= \frac{16}{9\epsilon^2} + \frac{8\pi^2}{27} + \f O(\epsilon),
\\ \notag
\f F_{lA}^{g(2)} &= \frac{2}{3\epsilon^3} + \frac{10}{9\epsilon^2} + \frac{1}{\epsilon} \left ( - \frac{52}{27} - \frac{\pi^2}{3} \right ) - \frac{1616}{81} - \frac{5\pi^2}{9} - \frac{148}{9} \zeta_3 + \f O(\epsilon),
\\
\f F_{lF}^{g(2)} &= - \frac{2}{\epsilon} - \frac{67}{3} + 16 \zeta_3 + \f O(\epsilon).
\end{align}

\section{Calculation of $\w Z$}
\label{app: wZ0}
\begin{figure}
    \centering
    \includegraphics[width=\linewidth]{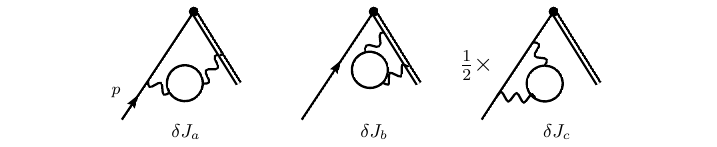}
    \caption{Leading non-vanishing loop graphs contributing to $\sqrt{\w Z}$. The first graph from the right $\delta J_c$ denotes the LSZ contribution.}
    \label{fig: deltaJ}
\end{figure}

The function $\w Z$ is defined by
\begin{align}
\sqrt{\w Z} u_c = \bra 0 W_c^{\dagger} \xi_c \ket{\hat p_c},
\end{align}
where $\hat p_c^{\mu} = n_+ p_c \frac{n_-^{\mu}}{2}$ denotes the massless collinear momentum with only the large light-cone component.
In particular, let $\delta J$ denote the sum of the graphs in fig. \ref{fig: deltaJ}. The longitudinal contribution $\propto k_{\mu} k_{\nu}$ in the photon self energy $\Pi_{\mu \nu}$ cancels between the three graphs. The remaining contributions are the transverse $\propto g_{\mu \nu}$ terms of $\delta J_a$ and $\delta J_c$, given by
\begin{align}
\delta J_a &=  - \frac{g_0^4 C_F}{(4\pi)^{d/2}} \int \frac{d^dk}{i \pi^{d/2}}  \frac{2  n_+ (k + p_c) \, \Pi(k^2, M_j^2) (w^2 \nu^{\eta} |n_+ k|^{-\eta}) }{(k^2 + n_- k \, n_+ p_c + i0)(k^2 + i0)(n_+ k + i0)},
\\
\delta J_c &=  - \frac{1}{2} \frac{g_0^4 C_F}{(4\pi)^{d/2}} \frac{d}{d\s p} \int \frac{d^dk}{i \pi^{d/2} k^2} \gamma_{\mu} \frac{1}{\s k + \s p} \gamma^{\mu} \Pi(k^2, M_j^2) \Big |_{\s p = 0}.
\end{align}
Let us start with $\delta J_a$.
Clearly, the $n_-k$ integral is only non-zero for $0 > n_+ k > - n_+ p_c$. We can perform the $n_-k$ integral by picking up the lower half plane pole at $n_-k = - \frac{k_{\perp}^2}{n_+(k+p_c)} - i0$. The remaining integral can be written as
\begin{align} \notag
\delta J_a &= - \frac{\alpha_0^2 }{\pi^2} \frac{ w^2  (4\pi)^{2\epsilon} \Gamma(\epsilon) }{\Gamma(1-\epsilon)} \left ( \frac{\nu}{Q} \right )^{\eta} C_F T_F \int_0^1 du \, u^{1-\epsilon} (1-u)^{1-\epsilon} \int_0^1 dx \, \frac{(1-x)^{1+\epsilon}}{x^{1+\eta} }
\\ \notag
&\quad \times \int_0^{\infty} \frac{ d|k_{\perp}^2| }{|k_{\perp}^2|^{1+\epsilon} } \, \frac{1}{ [ |k_{\perp}^2| + \frac{(1-x) M_j^2}{u(1-u)}  ]^{\epsilon}}
\\ \notag
&= \frac{1}{2} \left( \frac{\alpha_0}{4\pi} \right )^2  w^2(\nu) \left ( \frac{4\pi e^{-\gamma_E}}{M_j^2} \right )^{2\epsilon} \left ( \frac{\nu^2}{Q^2} \right )^{\eta/2} C_F T_F \frac{8}{3} \Bigg \{ \frac{1}{\eta} (\Gamma(\epsilon) e^{\epsilon \gamma_E})^2 \frac{1+\epsilon}{1+ \frac{8}{3} \epsilon + \frac{4}{3} \epsilon^2}
\\ \label{eq: delta Ja = 0}
&\quad + \frac{1}{\epsilon^2} + \frac{1}{\epsilon} \left ( - \frac{2}{3} - \frac{\pi^2}{6} \right ) + \frac{22}{9} + \frac{4\pi^2}{9} - \zeta_3 + \f O(\eta, \epsilon) \Bigg \},
\end{align}
where we have substituted $x = - \frac{n_+ k}{n_+p_c}$ in the intermediate result. The contribution from $\delta J_c$ is easily evaluated to be
\begin{align}
\delta J_c &= - \frac{1}{2} \left ( \frac{\alpha_0}{4\pi} \right )^2  \left ( \frac{4\pi e^{-\gamma_E}}{M_j^2} \right )^{2\epsilon}  C_F T_F \left \{ \frac{2}{3\epsilon^2} - \frac{13}{9\epsilon} + \frac{133}{54} + \frac{\pi^2}{9} + \f O(\epsilon) \right \}.
\end{align}
The two-loop contribution to $\w Z$ in eq. \eqref{eq: wZ0 twoloop} is then given by $2 \delta J_a + 2 \delta J_c$.

\newpage
\bibliographystyle{JHEP}
\bibliography{bibl}
\end{document}